\documentclass[aps,pra,twocolumn,amsmath,amssymb,nofootinbib,showpacs]{revtex4-1}
\usepackage[english]{babel}
\usepackage{times}
\usepackage{latexsym}
\usepackage{graphics}
\usepackage{epsfig}
\usepackage{color}
\usepackage{mathrsfs}
\usepackage[colorlinks]{hyperref}

\usepackage{color}
\definecolor{darkred}{rgb}{0.5,0,0}
\definecolor{darkgreen}{rgb}{0,0.5,0}
\definecolor{darkblue}{rgb}{0,0,0.5}

\hypersetup{colorlinks,
	linkcolor=blue,
	filecolor=blue,
	urlcolor=blue,
	citecolor=blue
}

\newcommand{\eg}{{e.g.\ }} % \ added to produce the correct spacing after the last dot

\renewcommand{\i}{\textrm{i}}
\newcommand{\ie}{{i.e., }}

\newcommand{\ft}[1]{\hat{#1}} % Fourier Transform
\newcommand{\ftadim}[1]{\ft{#1}}
\newcommand{\ftdim}[1]{#1}
\newcommand{\heaviside}{\Theta}
\newcommand{\mean}[1]{\overline{#1}}

\newcommand{\diffd}{d}
\newcommand{\expectationvalue}[1]{\langle #1 \rangle}

\newcommand{\bv}{\hat{b}}
\newcommand{\bvc}{\hat{b}^\dagger}

\newcommand{\ctwo}{c_2}
\newcommand{\ctwoFT}{\ftadim{c}_2}
\newcommand{\Ctwo}{C_2}
\newcommand{\CtwoFT}{\ftdim{C}_2}

\newcommand{\cthreeFT}{\ftadim{c}_3}
\newcommand{\Cthree}{C_3}
\newcommand{\CthreeFT}{\ftdim{C}_3}

\newcommand{\dens}{\hat{n}}
\newcommand{\Deltaone}{\Delta^{(1)}}
\newcommand{\Deltatwo}{\Delta^{(2)}}
\newcommand{\ddens}{\delta \hat{n}}

\newcommand{\epsk}{\epsilon_k}

\newcommand{\fm}{f^-}
\newcommand{\fp}{f^+}
\newcommand{\fpm}{f^\pm}

\newcommand{\ftmVk}{\ftdim{\modV}_k}
\newcommand{\fttVone}{\ftdim{\widetilde{V}}^{(1)}}

\newcommand{\gm}{g^-}
\newcommand{\gnBm}{g\mean{\nBEC}}
\newcommand{\gp}{g^+}
\newcommand{\gpm}{g^\pm}

\newcommand{\kmax}{k_{\mathrm{max}}}

\newcommand{\modC}{\mathcal{C}}
\newcommand{\modV}{\mathcal{V}}

\newcommand{\mVk}{\modV_k}
\newcommand{\mVm}{\modV_{-}}
\newcommand{\mVn}{\modV_n}

\newcommand{\nBEC}{n_\textrm{c}}

\newcommand{\phas}{\hat{\theta}}

\newcommand{\sigmar}{\sigma_\textrm{\tiny R}}

\newcommand{\tV}{\widetilde{V}}
\newcommand{\tVone}{\widetilde{V}^{(1)}}
\newcommand{\tVoner}{\tVone_\textrm{\tiny R}}
\newcommand{\tVtwo}{\widetilde{V}^{(2)}}
\newcommand{\tVr}{\tV_\textrm{\tiny R}}

\newcommand{\vect}[1]{\mathbf{#1}}
\newcommand{\vectf}{F}
\newcommand{\vectg}{G}
\newcommand{\vectr}{\vect{r}}
\newcommand{\vectrprime}{\vect{r^\prime}}

\newcommand{\vectq}{\vect{q}}
\newcommand{\Vr}{V_\textrm{\tiny R}}

\begin{document}

\title{Localization of Bogoliubov quasiparticles in interacting Bose gases with correlated disorder}

\author{P.~Lugan$^{1,2}$}
\author{L.~Sanchez-Palencia$^{1}$}
\affiliation{
$^1$Laboratoire Charles Fabry de l'Institut d'Optique, 
CNRS and Univ.~Paris-Sud,
Campus Polytechnique,
RD 128,
F-91127 Palaiseau cedex, France \\
$^2$Physikalisches Institut,
Albert-Ludwigs-Universit\"at,
Hermann-Herder-Str. 3, D-79104 Freiburg, Germany}
%\homepage{http://www.atomoptic.fr}

\date{\today}

\begin{abstract}
We study the Anderson localization of Bogoliubov quasiparticles (elementary many-body excitations) in a weakly interacting Bose gas of chemical potential $\mu$ subjected to a disordered potential $V$. We introduce a general mapping (valid for weak inhomogeneous potentials in any dimension) of the Bogoliubov-de Gennes equations onto a single-particle Schr\"odinger-like equation with an effective potential. For disordered potentials, the Schr\"odinger-like equation accounts for the scattering and localization properties of the Bogoliubov quasiparticles. We derive analytically the localization lengths for correlated disordered potentials in the one-dimensional geometry. Our approach relies on a perturbative expansion in $V/\mu$, which we develop up to third order, and we discuss the impact of the various perturbation orders.
Our predictions are shown to be in very good agreement with direct numerical calculations. We identify different localization regimes: For low energy, the effective disordered potential exhibits a strong screening by the quasicondensate density background, and localization is suppressed. For high-energy excitations, the effective disordered potential reduces to the bare disordered potential, and the localization properties of quasiparticles are the same as for free particles. The maximum of localization is found at intermediate energy when the quasicondensate healing length is of the order of the disorder correlation length. Possible extensions of our work to higher dimensions are also discussed.
\end{abstract}

\pacs{05.30.Jp, 03.75.Hh, 64.60.Cn, 79.60.Ht}

\maketitle

%%%%%%%%%%%%%%%%%%%%%%%%%%%%%%%%%%%%%%%%%%%
\section{Introduction}
\label{sec:introduction}

Disorder in ultracold quantum gases is attracting a growing interest due to
unprecedented possibilities of controlling disorder and atom-atom interactions
for bosons, fermions or mixtures of atomic species
created in one-, two- or three-dimensional traps~\cite{lsp2010,modugno2010,aspect2009,fallani2008}.
So far much attention has been devoted to studies of the disorder-induced damping of motion in Bose~\cite{lye2005,chen2008,albert2008,bhongale2010,dries2010,wu2011}
and Fermi~\cite{pezze2009} gases,
classical localization~\cite{clement2005,fort2005,schulte2005,clement2006,modugno2006,akkermans2008,lsp2008,mrsv2010,pezze2011b},
spatial diffusion~\cite{kuhn2005,lsp2005,shapiro2007,kuhn2007,beilin2010,mrsv2010,pezze2011b}
and Anderson localization in regimes where interactions can be neglected~\cite{damski2003,kuhn2005,lsp2007,lsp2008,kuhn2007,skipetrov2008,billy2008,roati2008,chabe2008,gurevich2009,lugan2009,lemarie2010,antezza2010,pezze2011a,piraud2011a,piraud2011b,moratti2011}.
The effects of disorder in interacting quantum systems have also been studied in a variety of contexts,
such as transport in weakly interacting Bose-Einstein condensates~\cite{paul2007,pikovsky2008,kopidakis2008,palpacelli2008,paul2009,flach2009,albert2010,hu2010,lucioni2011},
interacting Bose gases at equilibrium~\cite{damski2003,roth2003,gurarie2003,bilas2006,yukalov2007a,lugan2007a,lugan2007b,fallani2007,horstmann2007,yukalov2007b,gurarie2008,roux2008,nattermann2008,gaul2008,yukalov2009,falco2009,pollet2009a,gurarie2009,white2009,pasienski2010,deissler2010,aleiner2010,deissler2011,gaul2011a,gaul2011b},
strongly interacting Fermi gases~\cite{orso2007,byczuk2009,byczuk2010,han2011}
and coupled two-component gases~\cite{sanpera2004,ahufinger2005,wehr2006,niederberger2008,niederberger2009,niederberger2010,crepin2010}.

The interplay of disorder and interactions in quantum systems is an issue of fundamental importance to understand the behaviors of
superfluid $^4$He in porous media~\cite{bretz1973,reppy1984,reppy1992,crowell1995,crowell1997},
dirty superconductors~\cite{abrikosov1958,abrikosov1959,anderson1959,lee1985,belitz1994}
and ultracold gases in optical disorder~\cite{fallani2007,white2009,pasienski2010,deissler2010,deissler2011}.
Although a number of questions are open, in particular regarding the fate of Anderson localization,
general behaviors can be found in various situations.
For instance, weak repulsive interactions in a Bose gas at zero temperature
in a disordered potential generally lead to delocalization~\cite{lee1990,lugan2007a,nattermann2008,falco2009}:
In the absence of interactions, all bosons condense into the single-particle ground state,
which is localized~\cite{abrahams1979}.
This kind of $N$-body Fock state is highly unstable in an infinite system where an infinity of
spatially-separated single-particle states coexist at arbitrarily close energies~\cite{lifshits1988}.
Then, for weak repulsive interactions, the Bose gas fragments into a number of low-energy, localized single-particle states,
so as to minimize the interaction energy~\cite{lugan2007a,nattermann2008,falco2009}.
The Bose gas forms a gapless compressible insulating phase,
known as the Bose glass~\cite{ma1986,giamarchi1987,giamarchi1988,fisher1989,lee1990,scalettar1991,krauth1991,huang1992}.
For increasing mean-field interactions, the fragments merge and form a single extended condensate,
which restores superfluidity~\cite{lsp2006,lugan2007a,nattermann2008,falco2009,fontanesi2010,altman2010,fontanesi2011,vosk2011}.
Finally, in the strongly interacting regime, repulsive interactions can finally destroy again superfluidity, forming Tonks-Girardeau gases in 1D~\cite{hertz1979,giamarchi1988,demartino2005,radic2010} or disordered Mott insulators in lattice gases~\cite{fisher1989,scalettar1991,krauth1991}.

The above results lead to the conclusion that, at zero temperature where only the ground state plays a role,
repulsive interactions destroy Anderson localization in Bose gases
for moderate interaction strengths that are compatible with the mean-field approach~\cite{lee1990,lsp2006,lugan2007a}.
At non-zero temperature however, important properties,
such as the correlation functions, phase coherence and long-range order~\cite{fontanesi2009,fontanesi2010},
are determined by the excitations of the Bose gas, which are populated thermally.
It is thus of prime importance to determine how disorder affects the behavior of the latter.
These excitations of the many-body system are of collective nature. They can be viewed as quasiparticles,
scattering on the disordered potential.
In contrast to the mean-field background, which is extended,
the Bogoliubov quasiparticles of a weakly interacting Bose gas
can be localized in the presence of a disordered potential~\cite{bilas2006,lugan2007b,gurarie2008,gaul2011a,gaul2011b}.

In this paper, following the approach of Ref.~\cite{lugan2007b}, we present a detailed theory of the Anderson localization
of Bogoliubov quasiparticles in weakly interacting Bose gases subjected to correlated disordered potentials.
On the one hand, this approach applies to any kind of weak disordered potentials with short or long-range correlations
and offers a unified theory valid all along the crossover from the phonon regime to the free particle regime.
On the other hand, it permits a straightforward interpretation of the effect of repulsive interactions in terms of a screening of
the disorder by the density background.
In Sec.~\ref{sec:bogo}, the grand-canonical Hamiltonian of the interacting many-body system is expanded up to second order
in phase and density fluctuations.
The reduced Hamiltonian is quadratic and can be diagonalized by following the standard Bogoliubov-Popov approach.
The excitations are the solutions of the Bogoliubov-de Gennes equations, which consist in a set of two
coupled differential equations of order two.
In Sec.~\ref{sec:effectiveModel}, we then introduce a general mapping,
valid for weak inhomogeneous potentials in any dimension,
of the Bogoliubov-de Gennes equations onto a single-particle Schr\"odinger-like equation with a
screened potential.
For disordered potentials, the effective Schr\"odinger-like equation describes the scattering and localization properties
of the Bogoliubov quasiparticles.
In Sec.~\ref{sec:phaseFormBQPs}, we apply this approach to study the Anderson localization of Bogoliubov quasiparticles
in one-dimensional disorder.
We derive analytical formulas for the localization length of the Bogoliubov quasiparticles up to third order in perturbation
theory and compare our predictions to direct numerical calculations.
Our results exhibit different localization regimes:
For low energy, the effective disordered potential accounts for a strong screening by the quasicondensate density background and
Anderson localization is suppressed.
For high energy excitations, the screening is small;
the effective disordered potential reduces to the bare disordered potential
and the localization properties of quasiparticles are the same as for free particles.
The maximum of localization is found at intermediate energy when the quasicondensate healing length is of the order
of the disorder correlation length.
Finally, in Sec.~\ref{sec:conclusion}, we summarize our results and
discuss possible extensions of our work,
in particular towards higher dimensions.

%%%%%%%%%%%%%%%%%%%%%%%%%%%%%%%%%%%%%%%%%%%
\section{Elementary (Bogoliubov) excitations in a Bose gas with weak density fluctuations}
\label{sec:bogo}

We consider a $d$-dimensional, ultracold, dilute gas of bosons with weak repulsive interactions, in a potential $V(\vectr)$. The system is described by the grand-canonical Hamiltonian
%+++++++++++++++++++++++++++++++++++++++++%
\begin{eqnarray}
\hat{K}&=&\hat{H}-\mu\hat{N}\nonumber\\
& =& \int \diffd \vectr 
\left\{ \frac{\hbar^2}{2m} \left[(\vect{\nabla} \phas)^2 \dens
                     + (\vect{\nabla} \sqrt{\dens})^2\right] \right. \phantom{i}
\nonumber \\
         &&\left.  + V(\vectr) \dens + \frac{g}{2} \dens^2 - \mu \dens 
\right\},
\label{eq:Hamiltonian}
\end{eqnarray}
%+++++++++++++++++++++++++++++++++++++++++%
where the short-range atom-atom interactions are modeled by a contact potential with coupling constant $g > 0$, $m$ is the atomic mass, $\dens$ and $\phas$ are the density and phase operators, which satisfy the commutation relation $[ \dens (\vectr), \phas (\vect{r'}) ] = \i \delta ( \vectr - \vect{r'} )$~\cite{barnett1997}, and $\mu$ is the chemical potential. In the full form~(\ref{eq:Hamiltonian}), solving $\hat{K}$ for eigenstates is difficult in general. Yet, for small density fluctuations around $\nBEC=\expectationvalue{\dens}$ (\ie for $|\ddens|\ll \nBEC$, where $|\ddens|$ is the typical value of $\ddens = \dens - \nBEC$ in the state of the system) and for small phase gradient ($\hbar^2|\nabla \phas|^2/2m \ll \mu$), the operator $\hat{K}$ can be expanded around the classical field $\nBEC$, $\nabla\phas=0$, following the Bogoliubov-Popov approach \cite{bogoliubov1947,bogoliubov1958,popov1972,popov1983,mora2003,castin2004}.\footnote{The definition of a phase operator requires special care. A suitable definition can be found in Ref.~\cite{mora2003}, where a lattice model is used for a rigorous formulation of the Bogoliubov-Popov theory for quasicondensates. The equations derived in this lattice model coincide with the continuous formulation of Eqs.~(\ref{eq:GPE}), (\ref{eq:degennes1}) and (\ref{eq:degennes2}) upon replacement of the coupling constant $g$ by an effective coupling constant that depends on the lattice spacing $l$, and converges to~$g$ in the limit $l\to 0$ in 1D~\cite{mora2003,castin2004}.} In the zeroth-order expansion the ground-state density profile is found by minimizing the grand-canonical energy functional associated with the Hamiltonian~(\ref{eq:Hamiltonian}) with respect to the variation of $\nBEC(\vectr)$. This yields the Gross-Pitaevskii equation (GPE)
%+++++++++++++++++++++++++++++++++++++++++%
\begin{equation}
\left[-\frac{\hbar^2}{2m}\vect{\nabla}^2 + V(\vectr) + g \nBEC (\vectr) -\mu \right]\sqrt{\nBEC(\vectr)} = 0.
\label{eq:GPE}
\end{equation}
%+++++++++++++++++++++++++++++++++++++++++%
Then, retaining only the leading terms in the expansion of the density fluctuations $\ddens$ and phase gradients $\nabla\phas$, the Hamiltonian~(\ref{eq:Hamiltonian}) is cast into the form $\hat{K}=E_0 + \sum_\nu \epsilon_\nu\ \bvc_\nu \bv_\nu$, where $\bvc_\nu$ and $\bv_\nu$ are the bosonic creation and annihilation operators of an excitation (Bogoliubov quasiparticle, BQP) of energy $\epsilon_\nu$.\footnote{We discard in the canonical form of $\hat{K}$ and in Eqs.~(\ref{eq:opphas}) and~(\ref{eq:opdens}) the contribution of the $\hat{P}$ and $\hat{Q}$ operators which arise in Bogoliubov approaches without particle number conservation~\cite{lewenstein1996,castin1998}, as these operators play no role in the elementary excitations of the Bose gas and vanish in number-conserving approaches. Note also that the orthogonal projection of the Bogoliubov modes, as used in conserving approaches~\cite{castin1998,mora2003}, does not alter the results presented here.} The phase and density operators are expanded as
%+++++++++++++++++++++++++++++++++++++++++%
\begin{eqnarray}
 \phas (\vectr) &=& \frac{-i}{2\sqrt{\nBEC(\vectr)}} \sum_\nu \left[ \fp_\nu (\vectr)\ \bv_\nu - H.c. \right]  
\label{eq:opphas} \\
 \ddens (\vectr) &=& \sqrt{\nBEC(\vectr)} \sum_\nu \left[ \fm_\nu (\vectr)\ \bv_\nu + H.c. \right],
\label{eq:opdens}
\end{eqnarray}
%+++++++++++++++++++++++++++++++++++++++++%
where the Bogoliubov wavefunctions $\fpm_\nu(\vectr)$ are solutions of the Bogoliubov-de Gennes equations (BdGEs)~\cite{degennes1995}
%+++++++++++++++++++++++++++++++++++++++++%
\begin{eqnarray}
 \left[-\frac{\hbar^2}{2m} \vect{\nabla}^2 + V(\vectr)
+ \phantom{1}g\nBEC(\vectr)
-\mu \right] \fp_\nu (\vectr)
   &=& \epsilon_\nu \fm_\nu (\vectr)~~~
\label{eq:degennes1} \\
 \left[-\frac{\hbar^2}{2m} \vect{\nabla}^2 + V (\vectr)
+ 3g\nBEC (\vectr)
-\mu \right] \fm_\nu (\vectr)
   &=& \epsilon_\nu \fp_\nu (\vectr)~~~
\label{eq:degennes2}
\end{eqnarray}
%+++++++++++++++++++++++++++++++++++++++++%
with the normalization 
%+++++++++++++++++++++++++++++++++++++++++%
\begin{equation}
 \int \diffd \vectr \left[ \fp_{\nu}(\vectr) \fm_{\nu '}{}^*(\vectr) + \fm_{\nu}(\vectr) \fp_{\nu '}{}^*(\vectr) \right] = 2\delta_{\nu,\nu '}.
\end{equation}
%+++++++++++++++++++++++++++++++++++++++++%
Equations~(\ref{eq:opphas}) and (\ref{eq:opdens}) reveal the simple physical meaning of the functions $\fp_\nu(\vectr)$ and $\fm_\nu(\vectr)$. Up to the factor~$\sqrt{\nBEC(\vectr)}$, they describe the spatial dependence of the phase and density fluctuations associated with the BQPs, respectively. Notice that, as first pointed out by Popov~\cite{popov1972,popov1983}, the above derivation of the BdGEs in the phase-density representation provides an extension of the usual Bogoliubov-de Gennes theory. In the latter, the starting point consists in applying the usual Bogoliubov shift to the field operator, $\hat{\Psi}=\sqrt{\nBEC}+\delta\hat{\Psi}$, and expanding $\hat{K}$ up to quadratic terms in the fluctuation $\delta\hat{\Psi}$
\cite{bogoliubov1947,bogoliubov1958,degennes1995}. This approach assumes weak phase and density fluctuations around a unique classical field $\sqrt{\nBEC}$ which breaks the $U(1)$ phase symmetry of the Hamiltonian. On the contrary, the phase-density picture used in this work does not rely on this assumption. In particular, it provides a satisfactory description of the mean-field ground state $\nBEC({\vectr})$ and the excitations of Bose gases in the quasicondensate regime~\cite{petrov2000a,petrov2000b}. 

Within the above formalism, Eqs.~(\ref{eq:GPE}), (\ref{eq:degennes1}) and (\ref{eq:degennes2}) form a closed set which describes non-interacting, bosonic quasiparticles. Interactions between these quasiparticles only arise with higher-order terms in the expansion of $\hat{K}$, which we neglect here. Therefore, to study the low-temperature properties of the Bose gas in the external potential $V(\vectr)$, we are left with the sole modes defined by the GPE~(\ref{eq:GPE}) and BdGEs~(\ref{eq:degennes1}) and~(\ref{eq:degennes2}). Still, the set of equations~(\ref{eq:GPE}), (\ref{eq:degennes1}) and (\ref{eq:degennes2}) remains difficult to solve in general, as the GPE~(\ref{eq:GPE}) is nonlinear, and the two coupled second-order BdGEs~(\ref{eq:degennes1}) and (\ref{eq:degennes2}) themselves amount to a differential problem of order four. In the following, we develop a perturbative approach, valid in the limit of a weak potential $V(\vectr)$, which enables us to solve Eqs.~(\ref{eq:GPE}), (\ref{eq:degennes1}) and (\ref{eq:degennes2}) rigorously, and to interpret the underlying physics in simple terms.

%%%%%%%%%%%%%%%%%%%%%%%%%%%%%%%%%%%%%%%%%%%
\section{A Schr\"{o}dinger-like equation for Bogoliubov excitations in weak potentials}
\label{sec:effectiveModel}

From now on, we assume that $V(\vectr)$ is a weak external potential with a vanishing average ($\mean{V}=0$),\footnote{When $V$ is a disordered potential, we assume \textit{spatial homogeneity}, so that the spatial average of $V$ coincides with its statistical average~\cite{lifshits1988}. For non-disordered potentials, $\mean{V}$ denotes the spatial average of $V$.} and a typical amplitude $\Vr$,\footnote{The sign of $\Vr$ becomes relevant in the description of disordered potentials with asymmetric single-point probability distribution (see \eg Ref~\cite{lsp2010}).} such that $\Vr^2=\mean{V^2}$. While a less stringent weakness criterion can be derived (see below and Ref.~\cite{lsp2006}), $\vert\Vr\vert \ll \mu$ is a sufficient assumption to start with. Note that, although we will focus on the case of a disordered potential in the following, the perturbative approach introduced here is general, and $V$ need not be a disordered potential.
In any case, we write the autocorrelation function of $V$, $\Ctwo(\vectr'-\vectr) = \mean{V(\vectr)V(\vectrprime)}$, in the dimensionless form
%+++++++++++++++++++++++++++++++++++++++++%
\begin{equation}
 \Ctwo(\vectr)=\Vr^2 \ctwo(\vectr/\sigmar), 
\label{eq:correlationFunction}
\end{equation}
%+++++++++++++++++++++++++++++++++++++++++%
where $\sigmar$ is a characteristic length scale of $V$, which will be precisely defined below when needed. In the following paragraphs, we solve the GPE~(\ref{eq:GPE}) for the ground-state density $\nBEC$ (Sec.~\ref{subsec:BECdensity}), and we use the result to reduce the BdGEs~(\ref{eq:degennes1}) and~(\ref{eq:degennes2}) to a single Schr\"{o}dinger-like equation (Secs.~\ref{subsec:decouplingBdGEs} and \ref{subsec:effectiveSchroedinger}).

%%%%%%%%%%%%%%%%%%%%%%%%%%%%%%%%%%%%%%%%%%%
\subsection{The (quasi-)BEC density background}
\label{subsec:BECdensity}

In the regime where the repulsive interactions are strong enough (while remaining compatible with the mean-field regime), \ie when the healing length
%+++++++++++++++++++++++++++++++++++++++++%
\begin{equation}
\xi=\frac{\hbar}{\sqrt{4m\mu}}
\label{eq:healing}
\end{equation}
%+++++++++++++++++++++++++++++++++++++++++%
is much smaller than the size of the system $L$, the density profile $\nBEC$ is homogeneous in the absence of an external potential, and remains extended (delocalized) for a weak potential~$V$, owing to the nonlinear term in the GPE~(\ref{eq:GPE})~\cite{lee1990,lsp2006,lugan2007a}. It then proves useful to write the mean-field density term in the form
%+++++++++++++++++++++++++++++++++++++++++%
\begin{equation}
\nBEC (\vectr) = \frac{\mu + \Delta - \tV (\vectr)}{g},
\label{eq:nBECexact}
\end{equation}
%+++++++++++++++++++++++++++++++++++++++++%
where
%+++++++++++++++++++++++++++++++++++++++++%
\begin{equation}
\tV (\vectr) = \gnBm - g\nBEC (\vectr)
\end{equation}
%+++++++++++++++++++++++++++++++++++++++++%
contains the inhomogeneous part such that $\mean{\tV}=0$, and
%+++++++++++++++++++++++++++++++++++++++++%
\begin{equation}
 \Delta = \gnBm - \mu
\end{equation}
%+++++++++++++++++++++++++++++++++++++++++%
represents the mean deviation from the mean-field equation of state $\mu=g\nBEC$ that holds in the homogeneous case ($V=0$). The quantities $\tV$ and $\Delta$ both vanish for $V=0$, and are expected to remain small for a weak external potential $V$ and repulsive interactions that are strong enough. We then write the perturbation expansions of these quantities in increasing powers of $\Vr/\mu$:
%+++++++++++++++++++++++++++++++++++++++++%
\begin{eqnarray}
 \tV (\vectr) & = & \tVone (\vectr) + \tVtwo (\vectr) + \ldots \\
 \Delta & = & \Deltaone + \Deltatwo + \ldots.
\end{eqnarray}
%+++++++++++++++++++++++++++++++++++++++++%
The various terms can be calculated by generalizing the approach of Ref.~\cite{lsp2006} beyond the first order. Details of these calculations are presented in Appendix~\ref{sec:smoothing2} (see also Ref.~\cite{gaul2011b}). Below, we only discuss the main results.

%%%%%%%%%%%%%%%%%%%%%%%%%%%%%%%%%%%%%%%%%%%
\subsubsection{First correction to the mean-field equation of state}
\label{subsubsec:corrEqState}

\begin{figure}[t!]
\begin{center}
\includegraphics[width=8cm]{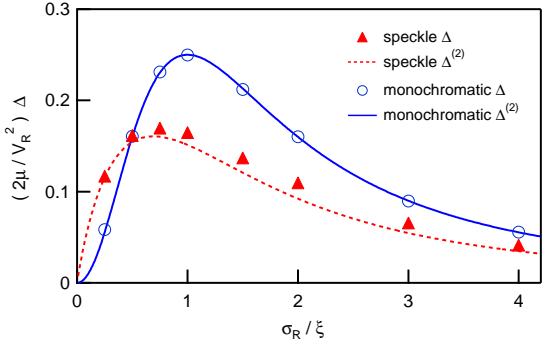}
\end{center}
\caption{(color online)
Comparison of the leading-order correction  $\Deltatwo$ to the mean-field equation of state with numerical computations of $\Delta$ for a 1D speckle potential with a reduced autocorrelation function $\ctwo(u)=\sin (u)^2/u^2$ as used in experiments~\cite{billy2008,clement2005,clement2006} and for a 1D periodic lattice  $V(z)=\Vr \sqrt{2} \mbox{ cos}(z/\sigmar)$. Here $\Vr=0.1\mu$.
}
\label{fig:delta}
\end{figure}

The first-order term $\Deltaone$ in the deviation $\Delta$ vanishes~\cite{lsp2006}. The leading term is thus provided by~$\Deltatwo$, which depends explicitly on the potential $V$ and on the healing length $\xi$ through [see Eq.~(\ref{eq:Deltatwobis}) in Appendix~\ref{sec:smoothing2}]: 
%+++++++++++++++++++++++++++++++++++++++++%
\begin{equation}
\label{eq:Deltatwo}
\Deltatwo = \frac{\Vr^2\sigmar^d}{2(2\pi)^{d/2}\mu}
            \int \diffd\vectq\ \frac{(|\vectq|\xi)^2}{\left[1+(|\vectq|\xi)^2\right]^2}\ctwoFT(\vectq\sigmar),
\end{equation}
%+++++++++++++++++++++++++++++++++++++++++%
where $\ctwoFT$ is the Fourier transform\footnote{Throughout the paper, the Fourier transform is defined as $f (\vect{q}) = (2\pi)^{-d/2} \int \diffd\vectr f(\vectr) e^{-i\vect{q} \cdot \vectr}$. The notation $\ftadim{f}$ is used for the Fourier transform of functions $f$ with dimensionless arguments.} of the reduced autocorrelation function $\ctwo$ defined in Eq.~(\ref{eq:correlationFunction}). As $\ctwoFT$ is a positive function by virtue of the Wiener-Khinchin theorem, we always have $\Deltatwo>0$, \ie $\mu<\gnBm$ in the presence of an external potential (see also Ref.~\cite{lee1990}). In Fig.~\ref{fig:delta}, Eq.~(\ref{eq:Deltatwo}) is compared to exact numerical calculations of $\Delta$ for a disordered potential and a monochromatic lattice potential, with various values of the ratio $\sigmar/\xi$, in a 1D geometry. As expected, the agreement is good for values of $\Vr/\mu$ as low as in Fig.~\ref{fig:delta}. We checked that the small discrepancy between $\Deltatwo$ and $\Delta$ in the disordered case is due to contributions of the order of $\Vr^3$, which are absent in a monochromatic lattice. This validates the perturbative approach.

%-----------------------------------------%
%-----------------------------------------%

Let us examine some limiting cases. In the Thomas-Fermi limit, $\xi \ll \sigmar$, we find
%+++++++++++++++++++++++++++++++++++++++++%
\begin{equation}
\label{eq:DeltatwoTF}
\Deltatwo \simeq \frac{\Vr^2}{2(2\pi)^{d/2}\mu} \left(\frac{\xi}{\sigmar}\right)^2
                 \int \diffd\boldsymbol{\kappa}\ |\vect{\boldsymbol{\kappa}}|^2 \ctwoFT(\boldsymbol{\kappa}),
\end{equation}
%+++++++++++++++++++++++++++++++++++++++++%
so that $\Deltatwo/\mu \propto ({\Vr}/{\mu})^2 ({\xi}/{\sigmar})^2 \ll ({\Vr}/{\mu})^2$.
The opposite limit, $\xi \gg \sigmar$, corresponds in principle to the white-noise limit
which is obtained by letting the ratio $\sigmar/\xi$ vanish while keeping the product
$\Vr^2\sigmar^d$ constant. Then, the Fourier transform of the reduced autocorrelation function may be approximated by a constant, $\ctwoFT(\vect{\boldsymbol{\kappa}}) \simeq \ctwoFT(\vect{0})$. In fact, we find that the white-noise limit of expression~(\ref{eq:Deltatwo}) is correctly defined only in 1D, for which we obtain
%+++++++++++++++++++++++++++++++++++++++++%
\begin{equation}
\label{eq:DeltatwoDeltacorr}
\Deltatwo \simeq \frac{\sqrt{\pi}\Vr^2\sigmar}{4\sqrt{2}\mu\xi}\ctwoFT(0),
\end{equation}
%+++++++++++++++++++++++++++++++++++++++++%
so that $\Deltatwo/\mu \propto ({\Vr}/{\mu})^2 ({\sigmar}/{\xi}) \ll ({\Vr}/{\mu})^2$. In dimension higher than one, this limit cannot be defined because the integrand
$\diffd\vect{q}\ (|\vect{q}|\xi)^2 / \left[ 1+(|\vect{q}|\xi)^2\right]^{2}$ scales as $q^{d-3}\diffd q$ for high momenta ($q \gg \xi^{-1}$). The integral in Eq.~(\ref{eq:Deltatwo}) would thus be plagued by an ultraviolet divergence for $d \geq 2$ and a constant $\ctwoFT(\vect{\boldsymbol{\kappa}})=\ctwoFT(\vect{0})$. In other words, for $d \geq 2$, the quantity $\Deltatwo$ depends crucially on the precise form of the reduced autocorrelation function $\ctwo$.

%%%%%%%%%%%%%%%%%%%%%%%%%%%%%%%%%%%%%%%%%%%
\subsubsection{Inhomogeneous part of the (quasi-)BEC density}
\label{subsubsec:fluctuatingPart}

In contrast to the mean deviation $\Delta$, the leading contribution to the inhomogeneous part of the density profile is provided by the first-order term, which reads
%+++++++++++++++++++++++++++++++++++++++++%
\begin{equation}
 \tVone (\vectr) = \int \diffd\vectr' G_{\xi} (\vectr-\vectr') V(\vectr'),
 \label{eq:tVoner}
\end{equation}
%+++++++++++++++++++++++++++++++++++++++++%
where $G_{\xi}$ is the Green function associated with the differential operator
$-\xi^{2}\nabla^{2}+1$
(see Ref.~\cite{lsp2006} and Eq.~(\ref{eq:tVonebis}) in Appendix~\ref{sec:smoothing2}).
In Fourier space we have
%+++++++++++++++++++++++++++++++++++++++++%
\begin{equation}
\ftdim{G}_{\xi} (\vect{q})=\frac{(2\pi)^{-d/2}}{1+(|\vect{q}|\xi)^2},
\label{eq:Gxi}
\end{equation}
%+++++++++++++++++++++++++++++++++++++++++%
so that
%+++++++++++++++++++++++++++++++++++++++++%
\begin{equation}
\fttVone (\vect{q}) = \frac{\ftdim{V}(\vect{q})}{1+(|\vect{q}|\xi)^2}.
\label{eq:tVone}
\end{equation}
%+++++++++++++++++++++++++++++++++++++++++%
The healing length $\xi$ clearly appears as a threshold length scale in the response of the density $\nBEC (\vectr)$ to the external potential $V(\vectr)$. Indeed, we have $\fttVone (\vect{q}) \simeq \ftdim{V}(\vect{q})$ for $|\vect{q}| \ll \xi^{-1}$, whereas $\fttVone (\vect{q}) \ll \ftdim{V}(\vect{q})$ for $|\vect{q}| \gg \xi^{-1}$. In other words, the potential $\tVone(\vectr)$ follows the spatial modulations of $V(\vectr)$ while evening out the high-frequency components. It also follows from the Parseval-Plancherel theorem that $\vert\tVoner\vert \leq \vert\Vr\vert$, where $\vert\tVoner\vert$ is the standard deviation of $\tVone$ and the sign of $\tVoner$ is chosen to be the same as that of $\Vr$. The potential $\tVone(\vectr)$ is thus termed a \textit{smoothed} potential~\cite{lsp2006}.

\begin{figure}[t!]
\begin{center}
\includegraphics[width=9cm]{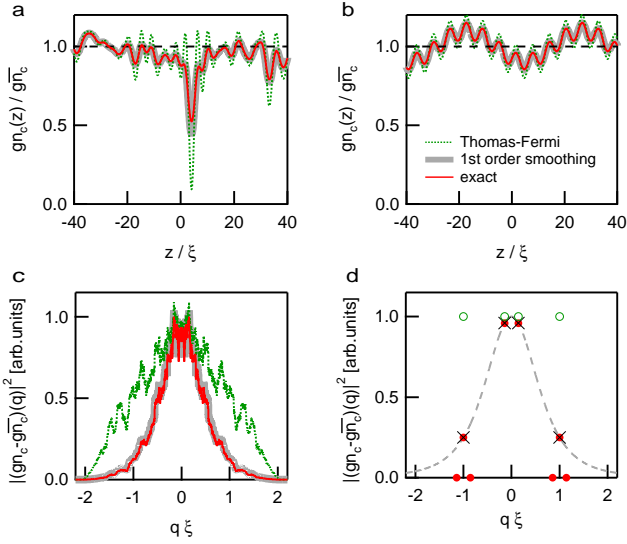}
\end{center}
\caption{(color online) Comparison of the first-order smoothing solution [Eq.~(\ref{eq:nBECexact}) with $\Delta=0$ and $\tV = \tVone$] and exact numerical computations of the density profile in the cases of (a,c) a 1D speckle potential with reduced autocorrelation function $c(u)=\sin (u)^2/u^2$ and correlation length $\sigmar=\xi$ and of (b,d) a 1D bichromatic periodic potential. The periodic potential reads $V(z)=\Vr[\mathrm{cos}(k_0 z)+\mathrm{cos}(k_0 z/7)]$ with $k_0 \xi = 1$. In both cases, $\Vr=0.1\mu$. (a,b) The analytical expression $\gnBm-\tVone(z)$ (broad grey line) is hardly distinguishable from $g\nBEC(z)$ as obtained numerically (red solid line). The average density $\gnBm$ (black dashed line) and the Thomas-Fermi limit $\gnBm-V(z)$ (green dotted line) are shown as references. (c,d) Corresponding power spectra of the modulations of $g\nBEC(z)$ (red solid line~/~red dots), $\tVone(z)$ (broad grey line~/~black crosses), and $V(z)$ (green dotted line~/~green circles). The grey dashed line in (d) is a plot of $1/[1+(q\xi)^2]^{2}$.
}
\label{fig:smoothing}
\end{figure}

If $V(\vectr)$ is a homogeneous disordered potential, that is, a disordered potential whose statistical properties do not depend on the position $\vectr$ \cite{lifshits1988}, then so is $\tVone(\vectr)$.
If $V(\vectr)$ is a periodic potential, $\tVone(\vectr)$ is also periodic with the same period, but a smoothed Bloch amplitude in each periodic cell, and simply rescaled Fourier components, as shown in Fig.~\ref{fig:smoothing}.
%-----------------------------------------%
%-----------------------------------------%
In either case, Eq.~(\ref{eq:nBECexact}) implies that, if $V(\vectr)$ is a homogeneous potential, the density profile $\nBEC$ is extended~\cite{lsp2006,lugan2007a}. Note also that the first-order term $\tVone$ is indeed a small perturbation of the homogeneous density profile whenever $\vert\tVoner\vert \ll \mu$, which loosens the initial weakness criterion $\vert\Vr\vert \ll \mu$.

The smoothing solution~(\ref{eq:nBECexact}) at first order ($\Delta=0$ and $\tV = \tVone$) is compared to exact numerical computations of the density profile in Fig.~\ref{fig:smoothing}, both for a disordered and a bichromatic periodic potential, in coordinate and Fourier spaces. For the sake of clarity, the power spectra for the disordered potential [Fig.~\ref{fig:smoothing}(c)] have been smoothed by a running average of width $0.1$ in $q\xi$ units. The agreement between the first-order smoothing solution and the numerical results is good, especially when compared to the Thomas-Fermi limit [for which $\tVone(z)=V(z)$]. For the bichromatic periodic potential, the appearance of red dots around $q\xi=\pm1$ in Fig.~\ref{fig:smoothing}(d) results from the admixture of the two components of the bichromatic lattice potential due to the nonlinearity of the GPE, as described by the first nonlinear correction $\tVtwo$ (see appendix~\ref{sec:smoothing2}). The other higher-order components, which are several orders of magnitude smaller, are not shown.

%%%%%%%%%%%%%%%%%%%%%%%%%%%%%%%%%%%%%%%%%%%
\subsection{Bogoliubov-de Gennes equations in the decoupling basis}
\label{subsec:decouplingBdGEs}

With the BEC background solution $\nBEC (\vectr)$ at hand, we can now solve the BdGEs~(\ref{eq:degennes1}) and (\ref{eq:degennes2}). Using Eq.~(\ref{eq:nBECexact}), these equations can be cast into
%+++++++++++++++++++++++++++++++++++++++++%
\begin{eqnarray}
\left[-\frac{\hbar^2}{2m} \vect{\nabla}^2 + V + \Delta - \tV \right] \fp_\nu 
   & = & \epsilon_\nu \fm_\nu
\label{eq:BdGEtV1} \\
\left[-\frac{\hbar^2}{2m} \vect{\nabla}^2 + 2\mu + V + 3\Delta - 3 \tV \right] \fm_\nu
   & = & \epsilon_\nu \fp_\nu,
\label{eq:BdGEtV2}
\end{eqnarray}
%+++++++++++++++++++++++++++++++++++++++++%
where $\Delta$, $V(\vectr)$ and $\tV(\vectr)$ are small compared to $\mu$.
As $\nBEC (\vectr)$ is the \textit{ground-state} solution of the GPE (\ref{eq:GPE}), the (quasi-) condensate is dynamically stable (see \eg\ Refs. \cite{castin2001,skryabin2000}), and we need only consider real-valued, strictly positive eigenvalues of the BdGEs. Now, given such an eigenvalue $\epsilon_{\nu}$, we are interested in the properties of the corresponding mode $\{\fp_{\nu}(\vectr),\fm_{\nu}(\vectr)\}$. Following the approach developed in Ref.~\cite{lugan2007b}, we take advantage of the structure of the eigenmodes before perturbation by a weak potential and introduce an alternative representation of the BQPs in terms of the components $\{\gp_{\nu}(\vectr),\gm_{\nu}(\vectr)\}$, where the functions $\gpm$ and $\fpm$ are related by the linear transformation
%+++++++++++++++++++++++++++++++++++++++++%
\begin{equation}
\label{eq:gpm}
g^\pm_{\nu} (\vectr) = \pm\rho_{\nu}^{\pm 1/2}\fp_{\nu}(\vectr) +\rho_{\nu}^{\mp 1/2} \fm_{\nu}(\vectr),
\end{equation}
%+++++++++++++++++++++++++++++++++++++++++%
with
%+++++++++++++++++++++++++++++++++++++++++%
\begin{equation}
\label{eq:rho}
\rho_{\nu} = \frac{\mu}{\epsilon_{\nu}}+\sqrt{1+\left(\frac{\mu}{\epsilon_{\nu}}\right)^2}.
\end{equation}
%+++++++++++++++++++++++++++++++++++++++++%
Expression~(\ref{eq:rho}) defines $\rho_\nu$ as a function of the eigenvalue $\epsilon_\nu$ which does not depend on the details of the specific mode under consideration. The linear transformation~(\ref{eq:gpm}), derived in Appendix~\ref{sec:decouplingBasis}, is chosen is such a way that the coupling between $\gp_{\nu}(\vectr)$ and $\gm_{\nu}(\vectr)$ vanishes in the homogeneous case $V=0$. As such, it offers a convenient starting point for a perturbation expansion in the case of weak potentials, as shown below.

In the basis of the $\gpm_\nu(\vectr)$ functions, the BdGEs take the exact form [see Eqs.~(\ref{eq:decoupled1Appendix})-(\ref{eq:decoupled2Appendix}) in Appendix~\ref{sec:decouplingBasis}]:
%+++++++++++++++++++++++++++++++++++++++++%
\begin{eqnarray}
\frac{\hbar^2 k_{\nu}^2}{2m} \gp_{\nu}
& = &
- \frac{\hbar^2}{2m} \nabla^2 \gp_{\nu}
- \frac{2\rho_{\nu}}{1+\rho_{\nu}^2} (\tV-\Delta) \gm_{\nu}
\nonumber \\
&& + \left[V-\frac{3+\rho_{\nu}^2}{1+\rho_{\nu}^2} (\tV-\Delta)\right] \gp_{\nu}
\label{eq:decoupled1} \\
- \frac{\hbar^2 \beta_{\nu}^2}{2m} \gm_{\nu}
& = &
- \frac{\hbar^2}{2m} \nabla^2 \gm_{\nu}
- \frac{2\rho_{\nu}}{1+\rho_{\nu}^2} (\tV-\Delta) \gp_{\nu}
\nonumber \\
&&  + \left[V-\frac{1+3\rho_{\nu}^2}{1+\rho_{\nu}^2} (\tV-\Delta)\right] \gm_{\nu},
\label{eq:decoupled2}
\end{eqnarray}
%+++++++++++++++++++++++++++++++++++++++++%
where
%+++++++++++++++++++++++++++++++++++++++++%
\begin{eqnarray}
 \frac{\hbar^2 k_{\nu}^2}{2m}
& = &
\sqrt{\mu^2+\epsilon_{\nu}^2} -\mu
\label{eq:kappa} \\
\frac{\hbar^2 \beta_{\nu}^2}{2m}
& = &
\sqrt{\mu^2+\epsilon_{\nu}^2} + \mu.
\label{eq:beta}
\end{eqnarray}
%+++++++++++++++++++++++++++++++++++++++++%
Both $k$ and $\beta$ are real-valued functions of the energy $\epsilon$. As a consequence, the associated $\gp_\nu$ and $\gm_\nu$ functions are essentially of the oscillating and the evanescent type, respectively, owing to the signs of the l.h.s.\ terms in Eqs.~(\ref{eq:decoupled1}) and (\ref{eq:decoupled2}). This is consistent with the limit of a vanishing external potential ($V=0$, and thus $\tV=0$, $\Delta=0$), where the equations for $\gp_\nu$ and $\gm_\nu$ are decoupled. It the latter case, the quantity $k$ can be identified with the wave number $|\vect{k}|$ of an oscillating, plane-wave BQP mode of energy $\epsilon_k$, and Eq.~(\ref{eq:kappa}) is equivalent to the usual Bogoliubov dispersion relation
%+++++++++++++++++++++++++++++++++++++++++%
\begin{equation}
\label{eq:BogoliubovDispersion}
\epsk=\sqrt{\frac{\hbar^2 k^2}{2m}\left(\frac{\hbar^2 k^2}{2m}+2\mu\right)}.
\end{equation}
%+++++++++++++++++++++++++++++++++++++++++%
The $\beta$ coefficients, on the other hand, characterize a subset of solutions to the BdGEs that are forbidden when $V=0$, as the boundary conditions imposed on the system (\eg periodic or homogeneous Dirichlet boundary conditions) preclude monotonously growing or decreasing BQP components.

%%%%%%%%%%%%%%%%%%%%%%%%%%%%%%%%%%%%%%%%%%%
\subsection{Effective Schr\"{o}dinger equation}
\label{subsec:effectiveSchroedinger}

\begin{figure}[t!]
\begin{center}
\includegraphics[width=8cm]{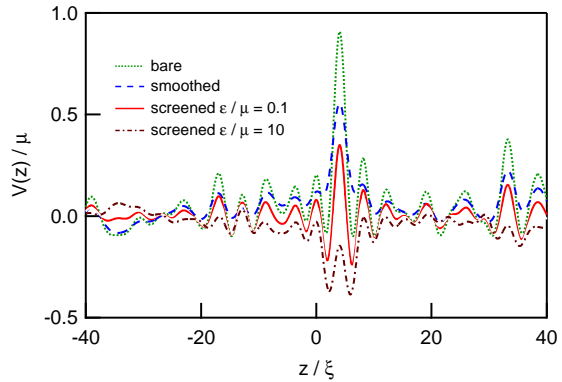}
\end{center}
\caption{(color online) Plot of the screened potential $\mVk(z)$ for the same 1D speckle potential ($\sigmar=\xi$, $\Vr/\mu=0.1$) as in Fig.~\ref{fig:smoothing}(a), with $\epsilon/\mu=0.1$ (\ie $k\xi\simeq0.05$) and $\epsilon/\mu=10$ (\ie $k\xi\simeq 2$). The bare potential $V(z)$ and the smoothed potential $\tV(z)$ are shown for comparison.
}
\label{fig:mVk}
\end{figure}

While the $\gm_{\nu}$ function vanishes identically in the absence of an external potential, this is no longer true when $V$ couples $\gm_{\nu}$ to $\gp_{\nu}$ via Eqs. (\ref{eq:decoupled1}) and (\ref{eq:decoupled2}). For a weak external potential, however, all the terms introduced by $\gm_{\nu}$ in Eq.~(\ref{eq:decoupled1}) are at least of second order in $\Vr$ [see Eq.~(\ref{eq:gkmConvolution}) and the discussion below], so that we can neglect the second term on the r.h.s.\ of Eq.~(\ref{eq:decoupled1}). Besides, the terms proportional to $\Delta$ in Eqs.~(\ref{eq:decoupled1}) and~(\ref{eq:decoupled2}) are at least of second order in the disorder amplitude~$\Vr$ (see Sec.~\ref{subsubsec:corrEqState}), and can also be neglected in a first-order approach. We are thus left with the following closed equation for $\gp_{\nu}$, which is valid to first order in $\Vr$:
%+++++++++++++++++++++++++++++++++++++++++%
\begin{equation}
-\frac{\hbar^2}{2m} \nabla^2 \gp_{\nu} (\vectr) +\modV_{k_\nu} (\vectr) \gp (\vectr)
\simeq \frac{\hbar^2 k_{\nu}^2}{2m} \gp_{\nu} (\vectr),
\label{eq:schroelike}
\end{equation}
%+++++++++++++++++++++++++++++++++++++++++%
where
%+++++++++++++++++++++++++++++++++++++++++%
\begin{equation}
\modV_{k_\nu}(\vectr) = V(\vectr) - \frac{3 + \rho_{\nu}^2}{1 + \rho_{\nu}^2} \tVone (\vectr).
\label{eq:effectiveV}
\end{equation}
%+++++++++++++++++++++++++++++++++++++++++%
Equation (\ref{eq:schroelike}) is formally equivalent to a Schr\"{o}dinger equation for a bare particle of energy $\hbar^2 k_{\nu}^2/2m$ in an effective potential $\modV_{k_\nu}(\vectr)$.\footnote{Note that the effective energy $\hbar^2 k_{\nu}^2/2m$ appearing in the Schr\"odinger-like equation~(\ref{eq:schroelike}) differs from the actual energy $\epsilon_{\nu}$ of the BQP under consideration, as shown by Eq.~(\ref{eq:kappa}).} The potential  $\modV_{k_\nu}(\vectr)$ differs from both the bare potential $V(\vectr)$ and the smoothed potential $\tV(\vectr)$ and explicitly depends on the BQP energy $\epsilon_{\nu}$ via the parameter $\rho_{\nu}$. These features are illustrated in Fig.~\ref{fig:mVk} where we plot a given realization of a 1D speckle potential $V(z)$, together with the corresponding smoothed potential $\tV(z)$ and effective potential $\modV_{k_\nu}$ (hereafter named \textit{screened} potential on grounds explained below), for two values of the BQP energy.

For further convenience, the dependence of the effective potential on the BQP energy $\epsilon_\nu$ is expressed by the subscript $k_\nu$, with the understanding that $k_\nu$ is defined by Eq.~(\ref{eq:kappa}). Combining Eqs.~(\ref{eq:rho}) and~(\ref{eq:kappa}), we find
%+++++++++++++++++++++++++++++++++++++++++%
\begin{equation}
\label{eq:rhok}
\rho_\nu=\sqrt{1+\frac{1}{(k_\nu\xi)^2}},
\end{equation}
%+++++++++++++++++++++++++++++++++++++++++%
and hence
%+++++++++++++++++++++++++++++++++++++++++%
\begin{equation}
\modV_{k_\nu}(\vectr) = V(\vectr) - \frac{1 + 4(k_\nu\xi)^2}{1 + 2(k_\nu\xi)^2} \tVone (\vectr).
\label{eq:effectiveVk}
\end{equation}
%+++++++++++++++++++++++++++++++++++++++++%
To gain more insight into the properties of $\modV_{k_\nu}(\vectr)$, let us turn to Fourier space where, by virtue of Eq.~(\ref{eq:tVone}), the effective potential reads
%+++++++++++++++++++++++++++++++++++++++++%
\begin{equation}
\modV_{k_\nu}(\vectq) = V(\vectq) \left[1 - \frac{1 + 4(k_\nu\xi)^2}{1 + 2(k_\nu\xi)^2} \frac{1}{1+(|\vect{q}|\xi)^2}\right].
\end{equation}
%+++++++++++++++++++++++++++++++++++++++++%
Upon inspection, this expression shows that we have $|\modV_{k_\nu}(\vectq)| \leq |V(\vectq)|$ for any Fourier component $\vectq$ and any BQP energy $\epsilon_\nu$.
Note also that, by construction of $\tVone$, the potential $\modV_{k_\nu}$ has a vanishing average. Hence, keeping in mind that $\modV_{k_\nu}$ results from the competition of the bare potential $V$ and the BEC background $\nBEC$ in the BdGEs, we term $\modV_{k_\nu}$ a {\it screened} potential. The screening thus affects all Fourier components of the external potential in any dimension.
%-----------------------------------------%
%-----------------------------------------%

Once Eq.~(\ref{eq:schroelike}) has been solved (possibly self-consistently) for $\gp_{\nu}$, the function $\gm_{\nu}$ can be computed from $\gp_{\nu}$.
Indeed, since $\hbar^2 \beta_{\nu}^2/{2m} > \mu$ [see Eq.~(\ref{eq:beta})]
and for $\vert\Vr\vert \ll \mu$, the last term in the r.h.s.\ of Eq.~(\ref{eq:decoupled2}) can be neglected, and we find
%+++++++++++++++++++++++++++++++++++++++++%
\begin{equation}
\label{eq:gkmConvolution0}
\gm_{\nu} (\vectr) \simeq
\frac{2m}{\hbar^2\beta_{\nu}^2}\frac{2\rho_{\nu}}{1+\rho_{\nu}^2} \int \diffd \vectr'
G_{1/\beta}(\vectr-\vectr') \tVone (\vectr') \gp_{\nu} (\vectr'),
\end{equation}
%+++++++++++++++++++++++++++++++++++++++++%
where $G_{1/\beta}(\vect{q})=\frac{(2\pi)^{-d/2}}{1+(|\vect{q}/\beta|)^2}$ is the Green function associated with the differential operator $-(1/\beta)^{2}\nabla^2+1$, written in Fourier space.

For analytical purposes, the $\gpm_{\nu}$ functions hence usefully replace the physically meaningful quantities $\fpm_{\nu}$, which can readily be recovered by inverting transformation (\ref{eq:gpm}). In particular, as far as asymptotic localization properties in disordered potentials are concerned, Eq.~(\ref{eq:gkmConvolution0}) tells us that the typical amplitude of $\gm_{\nu}$ evolves parallel to the amplitude of $\gp_{\nu}$ on intermediate to long length scales, if $\tV$ is homogeneous. In this respect, the benefit of the mapping of the exact BdGEs onto Eqs.~(\ref{eq:schroelike}) and (\ref{eq:effectiveV}) is that we can apply standard techniques for \textit{bare} Schr\"{o}dinger particles in weak disordered potentials, in any dimension, as long as these are consistent with the lowest-order approximation used to derive the effective equation~(\ref{eq:schroelike}). Yet, BQPs differ substantially from usual bare particles in their scattering and localization properties, because of the peculiar features of the screened potential $\modV_{k_\nu}(\vectr)$.

%%%%%%%%%%%%%%%%%%%%%%%%%%%%%%%%%%%%%%%%%%%
\section{Localization of Bogoliubov quasiparticles in one dimension}
\label{sec:phaseFormBQPs}

The formalism developed in Sec.~\ref{sec:effectiveModel} is valid for any weak potential. It is particularly fruitful when applied to disordered potentials as known theories developed for single (noninteracting) particles can be directly applied to the effective Schr\"odinger-like equation~(\ref{eq:schroelike}). In this section, we focus on the one-dimensional geometry using the so-called phase formalism~\cite{lifshits1988}, which allows for an exact perturbative calculation of the Lyapunov exponent (inverse localization length). This procedure can be straightforwardly extended to higher dimensions, applying appropriate single-particle theories to Eq.~(\ref{eq:schroelike}), for instance the self-consistent theory of localization~\cite{vollhardt1980a,vollhardt1980b}.

%%%%%%%%%%%%%%%%%%%%%%%%%%%%%%%%%%%%%%%%%%%
\subsection{Phase formalism in the Born approximation}

In 1D, the Lyapunov exponent $\gamma_k$ of a bare particle of energy $E_{k}$ in a disordered potential $V(z)$ is simply related to the backscattering amplitude of the particle from the inhomogeneities of $V$. For a weak disorder, the Lyapunov exponent can be extracted from a perturbation expansion $\gamma_k=\gamma_k^{(2)}+\gamma_k^{(3)}+\cdots$ in powers of the disorder amplitude $\Vr$.
For $\gamma_k \ll k$, this approach yields the following result in the lowest-order (Born) approximation~\cite{lifshits1988}:
%+++++++++++++++++++++++++++++++++++++++++%
\begin{equation}
\gamma_k^{(2)} = \frac{\sqrt{2\pi}}{8 k^2} \left(\frac{2m}{\hbar^2}\right)^2 \CtwoFT(2 k),
\label{eq:gammakBorn}
\end{equation}
%+++++++++++++++++++++++++++++++++++++++++%
where $\ftdim{C}_2(q)$ is the Fourier transform of $C_2(z)$, evaluated at $q=2k$. In this formulation, $V(z)$ has a vanishing average, and $k$, explicitly defined as $k=\sqrt{2mE_{k}}/\hbar$, stands for the typical wave vector of the particle under consideration. As such, the parameter $k$, rather than the related energy $E_k$, is the meaningful quantity in the interference effect that causes Anderson localization. Applying this result to the Schr\"{o}dinger-like equation~(\ref{eq:schroelike}), we derive the Lyapunov exponent $\Gamma_k$ of a BQP in the Born approximation $\Gamma_k \simeq \Gamma_k^{(2)}$:
%+++++++++++++++++++++++++++++++++++++++++%
\begin{equation}
\Gamma_k^{(2)} = \frac{\sqrt{2\pi}}{8 k^2} \left(\frac{2m}{\hbar^2}\right)^2 \ftdim{\modC}_{2,k}(2 k), 
\label{eq:Gk2}
\end{equation}
%+++++++++++++++++++++++++++++++++++++++++%
where $\ftdim{\modC}_{2,k}(q)$ is the Fourier transform of the two-point correlator $\modC_{2,k}(z)=\mean{\mVk(z^\prime)\mVk(z^\prime+z)}$, and $k$ depends on the energy $\epsilon$ through Eq.~(\ref{eq:kappa}) as we are now dealing with BQPs. From the Wiener-Khinchin theorem, we have $\ftdim{\modC}_{2,k}(q) \propto \mean{ |\ftmVk(q)|^2 }$, so that, according to Eq.~(\ref{eq:effectiveV}), the relevant Fourier component of $\mVk$ for the calculation of $\Gamma_k^{(2)}$ is
%+++++++++++++++++++++++++++++++++++++++++%
\begin{equation}
\ftmVk(2 k) =  V(2 k)-\frac{1+4(k\xi)^2}{1+2(k\xi)^2}\tVone(2 k).
\label{eq:Vscreen}
\end{equation}
%+++++++++++++++++++++++++++++++++++++++++%
Then, inserting Eq.~(\ref{eq:tVone}) into Eq.~(\ref{eq:Vscreen}), we obtain
%+++++++++++++++++++++++++++++++++++++++++%
\begin{equation}
\ftmVk(2k) =  \mathcal{S}(k\xi) \ftdim{V} (2k),
\label{eq:mV2k}
\end{equation}
%+++++++++++++++++++++++++++++++++++++++++%
where
%+++++++++++++++++++++++++++++++++++++++++%
\begin{equation}
\mathcal{S}(k\xi) = \frac{2(k\xi)^2}{1+2(k\xi)^2}.
\label{eq:Sfactor}
\end{equation}
%+++++++++++++++++++++++++++++++++++++++++%
Finally, Eq.~(\ref{eq:Gk2}) can be rewritten as
%+++++++++++++++++++++++++++++++++++++++++%
\begin{equation}
\Gamma_k^{(2)} =  [\mathcal{S}(k\xi)]^2 \gamma_k^{(2)},
\label{eq:Gk2gk2}
\end{equation}
%+++++++++++++++++++++++++++++++++++++++++%
where
%+++++++++++++++++++++++++++++++++++++++++%
\begin{equation}
\gamma_{k}^{(2)} = \frac{\sqrt{2\pi}}{32} \left(\frac{\Vr}{\mu}\right)^2 \frac{\sigmar}{k^2\xi^4} \ctwoFT(2 k \sigmar).
\label{eq:gammakExplicit}
\end{equation}
%+++++++++++++++++++++++++++++++++++++++++%
Equation (\ref{eq:Gk2gk2}), together with Eqs. (\ref{eq:kappa}), (\ref{eq:Sfactor}), and (\ref{eq:gammakExplicit}), completely determines the Lyapunov exponent of a BQP of energy~$\epsilon$ in a weak, correlated, 1D disordered potential.

Remarkably, Eq.~(\ref{eq:Gk2gk2}) shows that the Lyapunov exponent of a BQP can be simply related to that of a bare Schr\"{o}dinger particle with the same average wave vector $k$~\cite{lugan2007b}. The effects of interactions, in particular, are entirely absorbed in the term $\mathcal{S}(k\xi)$ defined in Eq.~(\ref{eq:Sfactor}), which we call {\it screening} function on the basis of the analysis presented in section~\ref{subsec:highE} below. This function is shown in Fig.~\ref{fig:S2} and displays two regimes which can be traced back to the nature of the elementary excitations of the interacting Bose gas in the absence of disorder.
%-----------------------------------------%
\begin{figure}[t!]
\begin{center}
\includegraphics[width=7cm]{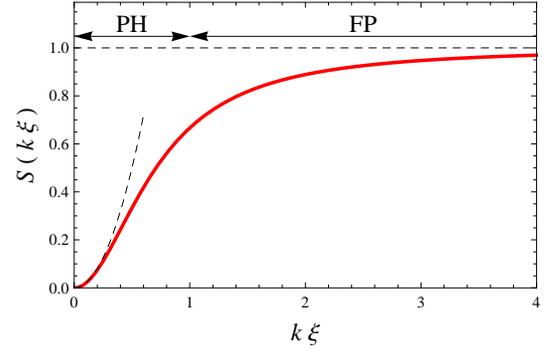}
\end{center}
\caption{Screening function $\mathcal{S}$. The dashed lines show the asymptotic behaviors in the phonon regime [$\mathcal{S}(k\xi) \simeq 2(k\xi)^2$ for $k\xi\ll 1$ or, equivalently, $\epsilon\ll \mu$] and free-particle regime [$\mathcal{S}(k\xi) \simeq 1$ for $k\xi\gg 1$ or, equivalently, $\epsilon\gg\mu$].
}
\label{fig:S2}
\end{figure}
%-----------------------------------------%
In the homogeneous case ($V=0$), the elementary (BQP) excitations of the Bose gas undergo a crossover from a regime of pair excitations with linear dispersion relation
for $\epsilon\ll\mu$, $\epsilon_k\simeq\hbar c k$ with $c=\sqrt{\mu/m}$ the speed of sound (phonon regime; PH), to a regime of single-particle excitations with a quadratic dispersion relation $\epsilon_k\simeq E_k=(\hbar^2/2m) k^2$ for $\epsilon\gg\mu$ (free-particle regime; FP). Hence, while the localization of single particles merely results from the competition of their kinetic energy with the disorder amplitude and correlation~\cite{lugan2009}, the localization of BQPs also crucially depends on interaction-induced particle correlations.

%%%%%%%%%%%%%%%%%%%%%%%%%%%%%%%%%%%%%%%%%%%
\subsection{Localization regimes}
\label{subsec:highE}

Summarizing the results of the previous section, we find that the Lyapunov exponent of BQPs in weak 1D disordered potentials is given by the expression
%+++++++++++++++++++++++++++++++++++++++++%
\begin{equation}
 \Gamma_k^{(2)} = \frac{\sqrt{2\pi}}{8}\left(\frac{\Vr}{\mu}\right)^2
\frac{k^2 \sigmar}{[1+2(k\xi)^2]^2} \ctwoFT(2k\sigmar),
\label{eq:Gamma0}
\end{equation}
%+++++++++++++++++++++++++++++++++++++++++%
obtained by combining Eqs.~(\ref{eq:Sfactor}),~(\ref{eq:Gk2gk2}) and~(\ref{eq:gammakExplicit}). 
In this expression, the quadratic dependence on the potential amplitude $\Vr$ is characteristic of the Born approximation. The scaled Lyapunov exponent $(\mu/\Vr)^2\Gamma_{k}^{(2)}\xi$ depends only on the two parameters $k\xi$ and $\sigmar/\xi$. Expression~(\ref{eq:Gamma0}) nevertheless contains contributions of three distinct physical origins, which appear more clearly in Eqs.~(\ref{eq:Gk2gk2}) and (\ref{eq:gammakExplicit}):
i)~a $1/k^2$ term which is representative of the kinetic energy of a bare particle,
ii)~the squared screening function $\mathcal{S}(k\xi)^2$ which accounts for particle interactions, and
iii)~the spectral density of disorder $\Vr^2\sigmar\ctwoFT(2k\sigmar)$ at the wave vector $2k$.
The role of these various contributions is discussed below.

%%%%%%%%%%%%%%%%%%%%%%%%%%%%%%%%%%%%%%%%%%%
\subsubsection{Screening in the phonon regime}
\label{subsubsec:screening}

The interplay of the first two contributions is best understood by studying the case of a white-noise potential, which is obtained in the limit $\sigmar\to 0$, $\vert\Vr\vert\to\infty$, with $\Vr^2\sigmar=\mathrm{const}$. In this limit, the spectral density $\ctwoFT$ uniformly approaches a flat distribution with an amplitude of the order of one. Then, in the free-particle regime $\epsilon\gg\mu$ (\ie $k\xi\gg1$), we have $\mathcal{S}(k\xi) \simeq 1$ and  $\Gamma_{k}^{(2)}\sim 1/k^2 \sim 1/\epsilon$. In other words, BQPs localize exactly like bare Schr\"{o}dinger particles in this regime, as expected. In the phonon regime, on the contrary, the kinetic term is dominated by the $\mathcal{S}(k\xi)^2$ factor, which is approximately quartic in $k$ (or $\epsilon$). We then get the scaling $\Gamma_{k}^{(2)}\sim k^2\sim\epsilon^2$~\cite{bilas2006,lugan2007b}, which is consistent with known results on the localization of acoustic phonons in 1D~\cite{ishii1973,jackle1981,john1983,azbel1983}.

Interestingly, Eq.~(\ref{eq:Gk2gk2}) combines the two limiting models in a unified picture, and provides a physical interpretation for the decreasing localization of phonon modes with decreasing energy. The function $\mathcal{S}(k\xi)$ reflects the competition of the bare external potential $V$ and the interaction of the BQPs with the quasi-BEC density background $g\nBEC$, which appears here as $\tV$. In particular, the strong decay of $\mathcal{S}(k\xi)$ in the phonon regime can be interpreted as an increasing screening of the external potential by the static quasi-BEC background, which adapts to the long-wavelength modulations of the disordered potential (see Sec.~\ref{subsubsec:fluctuatingPart}).

%%%%%%%%%%%%%%%%%%%%%%%%%%%%%%%%%%%%%%%%%%%
\subsubsection{Correlated disordered potentials}
\label{subsubsec:correlatedPotentials}

\begin{figure}[t!]
\begin{center}
\includegraphics[width=7.5cm]{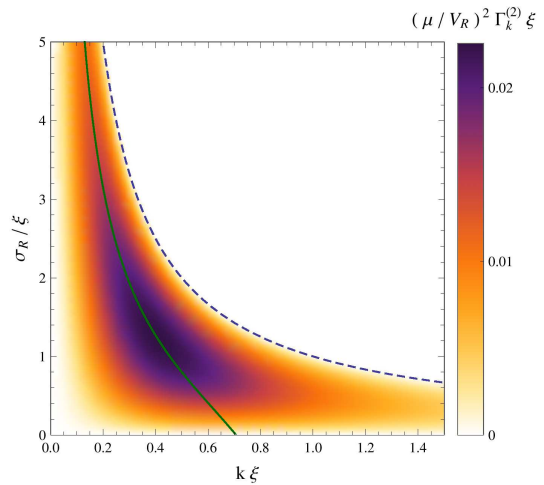}
\end{center}
\caption{Contour plot of the Lyapunov exponent of BQPs in a speckle potential,
as given by Eq.~(\ref{eq:Gamma0speckle}). Beyond $k\sigmar=1$ (black dashed line), the Lyapunov exponent vanishes completely in the Born approximation, due to the finite support of the speckle power spectrum. The green solid line represents the wave vector of maximum localization for each ratio $\sigmar/\xi$.}
\label{fig:GammaContour}
\end{figure}

To analyze the role of the correlation length $\sigmar$ in Eq. (\ref{eq:Gamma0}), we consider optical speckle potentials, which are now widely used with ultracold atoms for their tunability and truly random properties, as a model of correlated disorder~\cite{lye2005,clement2005,fort2005,clement2006,clement2008,chen2008,billy2008,white2009,dries2010,pasienski2010,mrsv2010}. In the simplest case where the speckle pattern is obtained at the back focal plane of a lens with rectangular aperture and uniform illumination (see e.g. Ref.~\cite{clement2006}), the reduced autocorrelation function reads
%+++++++++++++++++++++++++++++++++++++++++%
\begin{equation}
\ctwo(u)=
\sin(u)^{2}/u^{2},
\label{eq:speckleAutocorr}
\end{equation}
%+++++++++++++++++++++++++++++++++++++++++%
where $u=z/\sigmar$. The corresponding Fourier spectrum is
%+++++++++++++++++++++++++++++++++++++++++%
\begin{equation}
 \ctwoFT(k\sigmar)=
\sqrt{\frac{\pi}{2}}\left(1-\frac{k\sigmar}{2}\right)\heaviside\left(1-\frac{k\sigmar}{2}\right),
\label{eq:specklePowerSpectrum}
\end{equation}
%+++++++++++++++++++++++++++++++++++++++++%
where $\heaviside$ is the Heaviside step function. Then, Eq. (\ref{eq:Gamma0}) reads
%+++++++++++++++++++++++++++++++++++++++++%
\begin{equation}
\Gamma_k^{(2)} =
\frac{\pi}{8}\left(\frac{\Vr}{\mu}\right)^2
\frac{
	k^2 \sigmar (1-k\sigmar)
}{
	[1+2(k\xi)^2]^2
}
\heaviside(1-k\sigmar),
\label{eq:Gamma0speckle}
\end{equation}
%+++++++++++++++++++++++++++++++++++++++++%
which is shown in Fig.~\ref{fig:GammaContour}.
%-----------------------------------------%
%-----------------------------------------%
Such disorder correlations introduce several features, which we discuss below.

\paragraph*{Effective mobility edge -} Equation~(\ref{eq:Gamma0speckle}) shows that
the Lyapunov exponent in the Born approximation, $\Gamma_k^{(2)}$, vanishes identically for $k\sigmar>1$ (see also Figs.~\ref{fig:GammaContour} and \ref{fig:GammaAnalyticNumeric}). This feature originates from the special correlation properties of speckle potentials, the power spectrum of which has a high-momentum cutoff [see Eq.~(\ref{eq:specklePowerSpectrum})], \ie contains no $2k$ component able to backscatter a wave travelling with wavevector $k>1/\sigmar$ according to the elastic process $+k \rightarrow -k$ at the level of the Born approximation~\cite{lsp2007,lsp2008}. The Born approximation consists in truncating the perturbation series in powers of $\Vr$ used to derive Lyapunov exponents after the leading order, proportional to $\Vr^2$. In agreement with the understanding of localization in 1D, whereby single particles are localized (\ie $\gamma_k>0$) at all energies under fairly general assumptions~\cite{abrahams1979}, higher-order terms in the perturbation expansions of $\gamma_k$ and $\Gamma_k$ are expected to carry contributions  which do not all vanish identically for $k\sigmar>1$. However, the larger power dependence on the small parameter $\Vr/\mu$ makes these terms negligible in the limit of weak disorder. In this limit, the Lyapunov exponents thus experience a sharp crossover and drop by orders of magnitude when $k$ is varied through the value $1/\sigmar$. Such a crossover characterizes an \textit{effective} mobility edge which strongly affects localization properties in finite-size systems. As a matter of fact, the third-order contribution to $\Gamma_k$, proportional to $\Vr^3$, can be shown to vanish abruptly for momenta above the same cutoff at $1/\sigmar$
(see Sec.~\ref{subsec:beyondBorn}), so that corrections to Eq.~(\ref{eq:Gamma0speckle}) beyond that cutoff scale as $\Vr^4$ at least\footnote{For a complete discussion in the framework of the pure Schr\"{o}dinger particles in speckle potentials, see for instance Refs.~\cite{gurevich2009,lugan2009}.}. This behavior is specific to potentials with cutoffs in their Fourier-transformed correlation functions~\cite{izrailev1999,tessieri2002,lsp2007,gurevich2009,lugan2009}.

\paragraph*{Localization maxima -} In the white-noise limit, the BQPs localize best for $k\xi=1/\sqrt{2}$ (i.e. $\epsilon=\sqrt{3}\mu$), that is, in the cross-over region between the phonon and the free-particle regime~\cite{bilas2006}. This behavior results from the competition of bare kinetic energy and mean-field interactions via the screening effect, as discussed above. With correlations, however, the detailed statistical properties of the disorder play a role as well, and the wave vector of maximum localization $\kmax$ decreases with increasing correlation length $\sigmar$. More generally, it can be checked from Eq.~(\ref{eq:Gamma0}) that for any correlated disorder with monotonously decreasing power spectrum akin to model~(\ref{eq:specklePowerSpectrum}), the wave vector of maximum localization is shifted to lower values than the corresponding white-noise value. The locus of $\kmax$ as a function of the correlation and healing lengths is plotted in green solid line in Fig.~\ref{fig:GammaContour}. For each $\sigmar/\xi$ ratio, we indeed find a unique maximum $\kmax$ with
%+++++++++++++++++++++++++++++++++++++++++%
\begin{align}
\kmax\simeq &\frac{1}{\sqrt{2}\,\xi}\left(1-\frac{\sigmar/\xi}{2\sqrt{2}}\right), & \sigmar \ll \xi,\\
\kmax \simeq &\frac{2}{3\,\sigmar}, & \sigmar \gg \xi.
\end{align}
%+++++++++++++++++++++++++++++++++++++++++%
These asymptotic expressions show that $\kmax$ is controlled by the longest length scale in the problem. Finally, we find an absolute maximum at fixed $\xi$ for $\sigmar=\sqrt{3/2}\,\xi$ and $k\xi=1/\sqrt{6}$, which yields a localization length
%+++++++++++++++++++++++++++++++++++++++++%
\begin{equation}
 L_\textrm{max} (\xi) = \Gamma_\textrm{max}^{-1} (\xi) = \frac{512\sqrt{6}}{9\pi}\left(\frac{\mu}{\Vr}\right)^2 \xi.
 \label{eq:LmaxSpeckle}
\end{equation}
%+++++++++++++++++++++++++++++++++++++++++%
Current experiments with ultracold atoms implement disordered potentials with correlation lengths of the order of $\sigmar \simeq 0.25\mu$m \cite{billy2008,white2009,pasienski2010,mrsv2010}, which yields $L_\textrm{max} \simeq 230\mu$m for $\Vr = 0.2\mu$. Since this value can be of the order of or even smaller than the system size, we conclude that localization of BQPs in ultracold Bose gases is relevant for present-day experiments.

%%%%%%%%%%%%%%%%%%%%%%%%%%%%%%%%%%%%%%%%%%%
\subsection{Validity of the leading-order result}
\label{subsec:validity}

Before turning to some numerical tests, let us review the conditions of validity of the results discussed above. The Born approximation for BQPs, which yields $\Gamma_k \simeq  \Gamma_k^{(2)} = [\mathcal{S}(k\xi)]^2 \gamma_k^{(2)}$, requires
(i) the first-order smoothing solution
[Eq.~(\ref{eq:nBECexact}) with $\tV$ replaced by $\tVone$
as given by Eq.~(\ref{eq:tVoner})],
(ii) the first-order decoupling of the $\gp$ and $\gm$ modes that leads to Eq.~(\ref{eq:schroelike}), and
(iii) the Born approximation $\gamma_k\simeq\gamma_k^{(2)}$ to be valid.
The \textit{weak disorder} condition $\vert\tVr\vert \ll \mu$ alone ensures (i) and (ii). Note that this criterion of weak disorder appears less stringent on the amplitude $\Vr$ of the bare potential,
since smoothing reduces the amplitude of $\tV$ with respect to $V$. The regime of validity of the Lyapunov exponent derived for \textit{Schr\"odinger} particles in a weak-disorder expansion is in itself a subtle issue, as the successive terms in the perturbation series all depend on the disorder amplitude and the energy of the particle. The resulting asymptotic series is well-behaved in the high-energy limit. A precise inspection of the low-energy limit,
where the terms of the series blow up, is necessary to exhibit a rigorous criterion for the validity of a truncated perturbation expansion (see \eg\ Ref. \cite{derrida1984}). For single particles, $\gamma_k\ll k$ is usually retained \cite{lifshits1988}. In physical terms, the localization length should exceed the typical wavelength of the particle. This sets a $\Vr$-dependent lower bound on the single-particle energies for which the perturbative result is meaningful. Translating the above criterion to BQPs ($\Gamma_k \ll k$), we obtain
%+++++++++++++++++++++++++++++++++++++++++%
\begin{equation}
  \frac{\vert\Vr\vert}{\mu}\sqrt{\frac{\sigmar}{\xi}} \sqrt{\ctwoFT(2k\sigmar)}
\ll
  (k\xi)^{3/2}+\frac{1}{2(k\xi)^{1/2}}.
\label{eq:condValidity}
\end{equation}
%+++++++++++++++++++++++++++++++++++++++++%
This condition of validity resembles the corresponding one for Schr\"odinger particles:
$({\vert\Vr\vert}/{\mu})\sqrt{({\sigmar}/{\xi})} \sqrt{\ctwoFT(2k\sigmar)} \ll (k\xi)^{3/2}$. As expected, the two coincide in the FP regime ($k \gg 1/\xi$). However, they differ significantly in the PH regime ($k \ll 1/\xi$). Indeed, for free particles, perturbation theory always breaks down at low energy (\ie\ for $k \rightarrow 0$). Conversely, for BQPs in the PH regime, the strong screening of the disordered potential leads to a completely different condition: $({\vert\Vr\vert}/{\mu})\sqrt{({\sigmar}/{\xi})} \sqrt{\ctwoFT(2k\sigmar)} \ll 1/(k\xi)^{1/2}$.
The latter is always valid at low energy, with the assumption that $\ctwoFT(2k\sigmar)$ is of the order of unity at most. We thus find that the validity condition~(\ref{eq:condValidity}) is easily satisfied on the whole spectrum by a potential that is weak enough, \ie for $({\vert\Vr\vert}/{\mu})\sqrt{({\sigmar}/{\xi})} \ll 1$.

%%%%%%%%%%%%%%%%%%%%%%%%%%%%%%%%%%%%%%%%%%%
\subsection{Numerical calculations}
\label{subsec:numCalcs}

In order to test the accuracy of our perturbative approach, we performed numerical calculations of the Lyapunov exponent of BQPs in a 1D speckle potential, for various ratios $\sigmar/\xi$. The first step consisted in determining the ground-state solution $\nBEC$ of the GPE~(\ref{eq:GPE}), using a propagation scheme in imaginary time. As a precise determination of $\nBEC (z)$ is required for a correct calculation of the low-energy eigenmodes of the BdGEs~(\ref{eq:degennes1}) and (\ref{eq:degennes2}), we compared the result of this procedure with the smoothing expansion including up to ten perturbation orders (see appendix~\ref{sec:smoothing2}). The values of $\Delta$ computed with the two methods for $\sigmar=\sqrt{3/2}\,\xi$ agreed within a relative difference $\delta\Delta/\Delta$ of 0.3\% for $\Vr=0.05\mu$, while the r.m.s.\ difference of the computed density profiles typically amounted to a few $10^{-4}\Vr/g$. Homogeneous Dirichlet boundary conditions were imposed in the calculations, \ie both the density profile $\nBEC$ and the BQP components $\fp$ and $\fm$ were constrained to vanish at the system boundaries.\footnote{Periodic boundary conditions were found to complicate the analysis of the asymptotic localization properties of BQP modes, due to the periodic regrowths of the localized states in space.} We used system sizes of the order of a few $10^5$ healing lengths, so that the corrections to the equation of state $\mu=g\mean{\nBEC}$ solely due to kinetic terms at the boundaries of the system were negligible,\footnote{In a 1D box of length $L$, the relative correction to the equation of state associated with the Dirichlet boundary conditions scales as $\sqrt{2E_L/g\mean{\nBEC}} \simeq 2\xi/L$ for $g\mean{\nBEC}/E_L\gg1$, where $E_L=\hbar^2/2mL^2$.} in particular as compared to the corrections introduced by the disordered potential, described in section~\ref{subsubsec:corrEqState}.

In a second step, the profile $\nBEC(z)$ obtained as described above was included into the exact BdGEs~(\ref{eq:degennes1}) and (\ref{eq:degennes2}), and the Bogoliubov modes were computed by solving the associated discretized eigenvalue problem. Eigenvalues and eigenvectors of the resulting large, non-Hermitian band matrices were obtained for a limited set of target energies using standard ARPACK routines.\footnote{This approach was preferred over other traditional methods for boundary-value problems, such as shooting algorithms~\cite{NumericalRecipes}, since the propagation of initial data by a differential operator $\xi\partial_z$ diverges exponentially on the length scale of a few healing lengths due to the coupling into evanescent modes in the case of Bogoliubov excitations (see appendix~\ref{sec:decouplingBasis}).} The Lyapunov exponent $\Gamma_k$ was then obtained by computing estimators of
%+++++++++++++++++++++++++++++++++++++++++%
\begin{equation}
-\mathop{\lim}_{z\to\infty} \frac{\mean{\ln[r(z)/r(z_{0})]}}{|z-z_{0}|},
\label{eq:GammaEstimator}
\end{equation}
%+++++++++++++++++++++++++++++++++++++++++%
where $r=\sqrt{(\gp)^2+(\partial_{z}\gp/k)^2}$ defines the envelope of $\gp$ (see \eg\ Refs.~\cite{lifshits1988,lugan2009}), and the abscissa $z_{0}$ refers to the localization center of each eigenstate. In order to obtain accurate estimates of the average value of the logarithm of $r(z)$ at infinity for a wide range of $\sigmar/\xi$ and $k\xi$ parameters, the numerical calculations were carried out in a large box of size $L=3.2\times10^5\xi$ (\ie $10^5-10^6\sigmar$), and disorder averaging was performed over 200 randomly generated 1D speckle patterns.
We checked that, as expected, using $r(z)=|\gp(z)|$ in expression~(\ref{eq:GammaEstimator}) produces the same results but requires more extensive disorder averaging, owing to the divergence of $\ln |\gp(z)/\gp(z_0)|$ at the nodes of $\gp$. We also found that replacing $\gp$ by one of the functions $\fp$, $\fm$ or $\gm$ in the expression of $r$ leaves our numerical estimates of the Lyapunov exponent~(\ref{eq:GammaEstimator}) unchanged, within a relative difference of less than $5\%$ for all the parameters in this study.

\begin{figure}[t!]
\includegraphics[width=8cm]{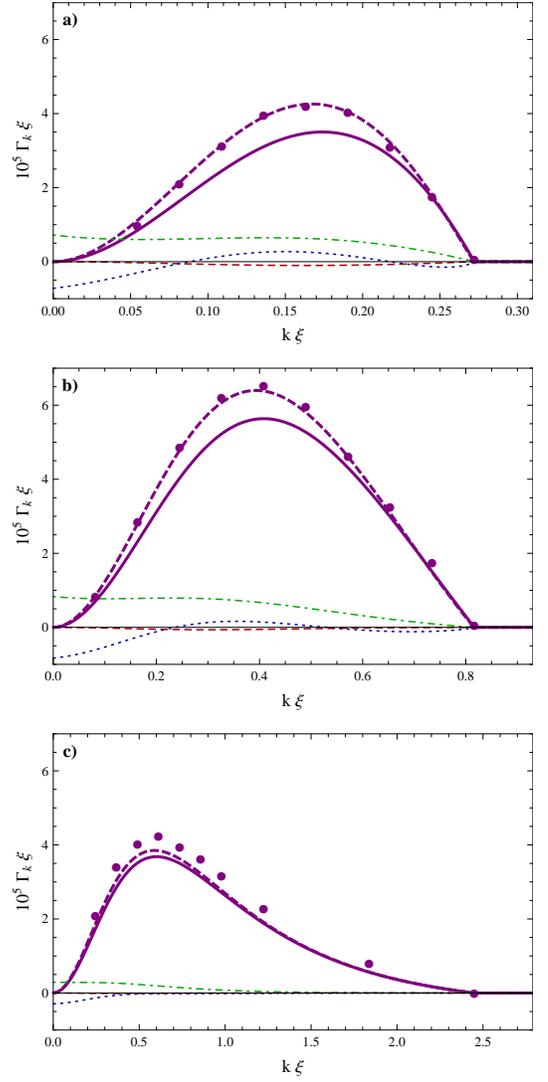}
\caption{
Lyapunov exponent of BQPs in
1D speckle potentials with the statistical properties described in section~\ref{subsubsec:correlatedPotentials}, with $\Vr/\mu=0.05$, and for a)~$\sigmar=3.7\,\xi$, b)~$\sigmar=\sqrt{3/2}\,\xi$, and c)~$\sigmar=0.4\,\xi$.
The thick solid lines correspond to the Born term $\Gamma_k^{(2)}$ given by Eq.~(\ref{eq:Gamma0speckle}). The thick purple dashed lines correspond to the next-order expansion $\Gamma_k\simeq\Gamma_k^{(2)}+\Gamma_k^{(3)}$  (see section~\ref{subsec:beyondBorn}). The term $\Gamma_k^{(3)}$ scales as $\Vr^3$, and results from the addition of three contributions: $\Gamma_{\mVk,\mVn}^{(3)}$ (green dash-dotted), $\Gamma_{\mVk,\mVm}^{(3)}$ (red dashed), and $\Gamma_{\mVk,\mVk,\mVk}^{(3)}$ (blue dotted). The dots are the numerical data obtained with the procedure
described in section~\ref{subsec:numCalcs}.
}
\label{fig:GammaAnalyticNumeric}
\end{figure}

On the whole spectral range which spans the phonon and free-particle regimes, the numerical data (filled points) shown in Fig.~\ref{fig:GammaAnalyticNumeric} are in excellent qualitative and fair quantitative agreement with the analytic prediction of Eq.~(\ref{eq:Gamma0speckle}) (thick solid lines), which corresponds to second-order perturbation theory. The choice of the value of $\Vr/\mu$ was motived by experimental relevance and, as aforesaid, by numerical tractability for an entire set of $\sigmar/\xi$ and $k\xi$ parameters.
 
%-----------------------------------------%
%-----------------------------------------%

Still, for this intermediate value of $\Vr/\mu$, we find slight deviations between the numerical data and Eq.~(\ref{eq:Gamma0speckle}). As discussed in Sec.~\ref{subsec:beyondBorn}, this small difference can be attributed mostly to third-order terms which contribute to the exact Lyapunov exponent $\Gamma_k$.
Hence, the observed deviations are expected to be negligible for lower values of $\Vr/\mu$.
We also find that, even if these deviations cannot be completely neglected for $\Vr \simeq 0.05\mu$,
they do not change the qualitative behavior of $\Gamma_k$.
Finally, note that for comparable parameters the deviations appear smaller in the data of Ref.~\cite{lugan2007b} compared to those of the present work. The present results are actually more accurate as only a lowest-order smoothing expansion was used to compute the density $\nBEC$ in Ref.~\cite{lugan2007b}.

%%%%%%%%%%%%%%%%%%%%%%%%%%%%%%%%%%%%%%%%%%%
\subsection{Beyond the Born approximation}
\label{subsec:beyondBorn}

While the screened potential $\mVk$ of Eq.~(\ref{eq:schroelike}) and the Lyapunov exponent $\Gamma_k^{(2)}$ of Eq.~(\ref{eq:Gk2gk2}) accurately account for the scattering and localization properties of BQPs in the limit of weak potentials, going beyond the leading-order (Born) approximation used to derive them is interesting in several respects. First, it should allow us to address the question of localization beyond cutoffs in the Lyapunov exponent $\Gamma_k^{(2)}$ such as the one arising in speckle potentials at wave number $1/\sigmar$, described in section~\ref{subsubsec:correlatedPotentials}. Second, studying the localization properties of BQPs for stronger disorder or weaker interactions is of particular importance, since with increasing $\Vr/\mu$ ratio the interacting Bose gas moves away from the deep superfluid (quasi-)BEC regime and into the weakly interacting fragmented Bose-glass phase~\cite{giamarchi1987,giamarchi1988,lugan2007a,fontanesi2009,fontanesi2010}. While recent studies~\cite{gurarie2008,fontanesi2009,fontanesi2010} have shown that the low-energy scaling of the inverse participation length of BQPs (which characterizes their short-range localization properties) is modified through the phase transition, the impact of strong disorder on the long-range localization properties of BQPs remains an open issue.

We now briefly address the localization properties of the Bogoliubov quasiparticles beyond the Born approximation by an inspection of the next-order terms of the weak-disorder expansion. While details of the derivation can be found in appendix~\ref{sec:appendixBeyondBorn}, here we just outline the approach. The starting point is again the set of BdGEs~(\ref{eq:decoupled1}) and~(\ref{eq:decoupled2}) in the decoupling basis $\gpm$. Retaining terms up to second order in the potential amplitude $\Vr$, we find a new approximate Schr\"odinger-like equation for $\gp$:
%+++++++++++++++++++++++++++++++++++++++++%
\begin{equation}
\frac{\hbar^2 k^2}{2m} \gp
\simeq
- \frac{\hbar^2}{2m} \nabla^2 \gp
+\left[\mVk(\vectr)+\mVn(\vectr)+\mVm(\vectr) \right] \gp,
\label{eq:decoupledSecondOrder}
\end{equation}
%+++++++++++++++++++++++++++++++++++++++++%
where $\mVk$ is the screened potential of Eq.~(\ref{eq:schroelike}), proportional to $\Vr$. The two additional terms $\mVn$ and $\mVm$ scale as $\Vr^2$, and account for the second-order modulations of the density profile $\nBEC$ and for the coupling of $\gp$ and $\gm$, respectively. Their expressions are given in Eqs.~(\ref{eq:mVn}) and (\ref{eq:mVm}).
Although Eq.~(\ref{eq:decoupledSecondOrder}) is valid {\it a priori}
only in the regime $k\ll \textrm{min}(1/\sigmar,1/\xi)$ (see Appendix~\ref{sec:appendixBeyondBorn}), we found it to provide a rather good approximation over a wider range of parameters (see Fig.~\ref{fig:GammaAnalyticNumeric} and below).
On the basis of the Schr\"odinger-like equation~(\ref{eq:decoupledSecondOrder}), we calculate the third-order contributions to the Lyapunov exponent following the approach of Ref.~\cite{lugan2009}. We find
%+++++++++++++++++++++++++++++++++++++++++%
\begin{equation}
\Gamma_k^{(3)}=\Gamma_{\mVk,\mVn}^{(3)}+\Gamma_{\mVk,\mVm}^{(3)}+\Gamma_{\mVk,\mVk,\mVk}^{(3)},
\label{eq:Gamma3contributions}
\end{equation}
%+++++++++++++++++++++++++++++++++++++++++%
where $\Gamma_{\mVk,\mVn}^{(3)}$ (resp. $\Gamma_{\mVk,\mVm}^{(3)}$) stems from the cross-correlator of $\mVk$ with $\mVn$ (resp. $\mVm$), and the last term involves the three-point autocorrelation function of $\mVk$ [see Eqs.~(\ref{eq:contrib2}) to (\ref{eq:contrib4})]. These various contributions are plotted in Fig.~\ref{fig:GammaAnalyticNumeric} for a speckle potential and various $\sigmar/\xi$ ratios. We find an excellent agreement between the numerical data and the analytical result $\Gamma_k^{(2)}+\Gamma_k^{(3)}$. For the smallest value of $\sigmar/\xi$ however, the third order term $\Gamma_k^{(3)}$ does not fully account for the small difference between the Born approximation $\Gamma_k^{(2)}$ and the numerical data [see Fig.~\ref{fig:GammaAnalyticNumeric}(c)]. This may be due to the fact that the criterion $k\ll 1/\xi$ is not met or to contributions of higher order in $\Gamma_k$. Let us now discuss the properties of the terms appearing in Eq.~(\ref{eq:Gamma3contributions}).

Note first that the third-order contributions scale as $\Vr^3$. They are thus non-zero only for potentials which do not possess symmetric statistics under the transformation $V\to -V$. This holds for speckle potentials as used in experiments with ultracold atoms~\cite{lugan2009}.

A generic feature of the contributions to $\Gamma_k^{(3)}$ is that they are all of the form
%+++++++++++++++++++++++++++++++++++++++++%
\begin{equation}
\Gamma_{i}^{(3)}\propto \int \diffd q F_i(q) \cthreeFT(q,2k\sigmar),
\end{equation}
%+++++++++++++++++++++++++++++++++++++++++%
where $i$ is an index labeling any of the terms in Eq.~(\ref{eq:Gamma3contributions}), $F_i$~is some function, and $\cthreeFT(q,q^\prime)$ is the Fourier transform of the reduced three-point autocorrelation function of the bare potential $V$. Thus, we find that  if $\cthreeFT$ has a compact support, $\Gamma_k^{(3)}$ vanishes over an extended part of the spectrum, as in the single-particle case~\cite{lugan2009,gurevich2009}. In particular, in the case of speckle potentials, one finds a high-momentum cutoff that is identical to the cutoff in $\ctwoFT$, whereby $\Gamma_k^{(3)}$ vanishes identically for $k>1/\sigmar$, just as $\Gamma_k^{(2)}$ does. This result is consistent with the argument that no odd power of $\Vr$ can provide the leading-order term in a given part of the spectrum, since the Lyapunov exponent is a non-negative quantity~\cite{lugan2009}. Note that this feature emerges from the analysis of the terms in Eq.~(\ref{eq:Gamma3contributions}) although the region $k\approx 1/\sigmar$ {\it a priori} lies outside the regime of validity of Eq.~(\ref{eq:decoupledSecondOrder}).

Figure~\ref{fig:GammaAnalyticNumeric} shows that the magnitude of $\Gamma_{\mVk,\mVm}^{(3)}$, which originates from the coupling of $\gp$ and $\gm$, is small compared to the other terms in $\Gamma_k^{(3)}$. This further legitimizes the use of the $\gpm$ basis and suggests that the difference between the analytical and the numerical results in Fig.~\ref{fig:GammaAnalyticNumeric}(c) are likely to be due to higher order terms rather than the violation of criterion $k\ll\xi$. We also note that, remarkably, the contributions $\Gamma_{\mVk,\mVn}^{(3)}$ and $\Gamma_{\mVk,\mVk,\mVk}^{(3)}$ tend to constant values at zero energy which turn out to be opposite
and thus cancel out in the calculation of $\Gamma_k^{(3)}$. This cancellation between two terms that seem to have different origins in the perturbation expansion is certainly not accidental, and must be due to the $\gpm$ representation chosen to set it up. Finally, an inspection of the low-energy limit of expression~(\ref{eq:Gamma3contributions}) in the case of the above speckle potential yields the scaling $\Gamma_k^{(3)}\sim k^2\sim E^2$, a feature which is likely to be generic for disordered potentials with $\cthreeFT(0,0)\neq0$.

%%%%%%%%%%%%%%%%%%%%%%%%%%%%%%%%%%%%%%%%%%%
\section{Conclusion}
\label{sec:conclusion}

In this work, we presented a general approach, valid in any dimension, to describe a weakly interacting Bose gas of chemical potential $\mu$ subjected to a weak inhomogeneous potential $V(\vect{r})$. This approach relies on a two-step perturbative expansion of the Gross-Pitaevskii equation (GPE) and the Bogoliubov-de Gennes equations (BdGE), which govern the (quasi-)condensate background and elementary excitations (Bogoliubov quasiparticles) of the Bose gas,
respectively. In the first step, we calculate the mean-field density profile using a perturbative expansion of the GPE in $V/\mu$. In the second step, the result is incorporated into the BdGEs. Turning to an adapted basis for the Bogoliubov wavefunctions, we then show that the BdGEs can be approximately mapped onto a Schr\"odinger-like equation, with an effective potential $\mVk$ which depends on the bare potential $V$, the condensate density background, and the quasiparticle wave vector~$k$.

Our approach is well suited to study the effects of disorder in interacting Bose gases
and, in particular, to examine the Anderson localization of Bogoliubov quasiparticles.
On the one hand, it applies to any kind of weak, correlated disordered potentials.
We stress that it is not limited to i) Gaussian disorder, ii) (uncorrelated) white-noise potentials or iii) models of non-overlapping impurities. In particular, it applies to speckle potentials as used in many experiments with ultracold atoms.
On the other hand, the only small parameter of the perturbative expansion is the ratio of the disorder amplitude $\Vr$ over the chemical potential $\mu$. Our approach differs in this respect from standard approximations used in various other works:
(i)~the approximation of the BdGEs by hydrodynamical equations, which confine the theory to excitations of typical wavelength $\lambda\gg\xi$, where $\xi$ is the healing length of the condensate~\cite{bilas2006,castin2001},
(ii)~the Thomas-Fermi approximation of the background density profile, which is valid only under the assumption $\xi\ll\sigmar$, where $\sigmar$ is the typical (minimal) length scale on which the external potential varies~\cite{bilas2006},
(iii)~the white-noise approximation which requires at least $\sigmar\ll\lambda$~\cite{bilas2006,giorgini1994}.
Conversely, the approach developed here holds for any ordering of the length scales $\xi$, $\sigmar$ and $\lambda$.

Although our approach can be used to describe the scattering and localization properties of Bogoliubov quasiparticles in any dimension, we focused in this work on the one-dimensional case, which leads to the strongest localization effects~\cite{lugan2007b}. 
In the low-energy limit and at the leading and next-leading orders in the disorder amplitude, we found a quadratic scaling of the Lyapunov exponent with the quasiparticle energy, $\Gamma_k \sim \epsilon^2 \propto k^2$. This finding is consistent with known results on the localization of phonons~\cite{ishii1973,jackle1981,john1983,azbel1983} and studies in the white-noise limit~\cite{bilas2006}. The effective potential $\mVk$ derived in our approach provides a physical interpretation of this suppression of localization in the low-energy phonon regime ($k\xi\ll1$) in terms of an efficient screening of the long-wavelength modulations of the external potential by the background density of the Bose gas. In the free-particle regime ($k\xi\gg1$), the Lyapunov exponent asymptotically approaches the exponent of a bare Schr\"odinger particle, $\Gamma_k \simeq \gamma_k$.
For uncorrelated potentials, the Lyapunov exponent of Bogoliubov quasiparticles thus falls off as $\Gamma_k \sim 1/\epsilon \propto 1/k^2$ in the high-energy limit.
For correlated disorder, the high-energy decay of the Lyapunov exponent strongly depends
on the large $k$ behavior of the disorder power spectrum,
$\Gamma_k \sim \CtwoFT(2k)/k^2$.
If $\CtwoFT(2k)$ has a finite support as for speckle potentials for instance~\cite{lsp2007,lsp2008,lugan2009}, effective mobility edges arise as for bare-particles~\cite{izrailev1999,tessieri2002,gurevich2009,lugan2009}.

Most importantly, our approach covers the crossover between the phonon and free particle regimes.
We find that localization ($\Gamma_k$) is maximum at a given energy $\epsilon$.
For uncorrelated potentials, this maximum lies around $E\simeq\mu$ (\ie $k\xi\simeq1$),
\ie at the crossover between the phonon and free-particle regimes~\cite{bilas2006,lugan2007b}.
For correlated potentials, the strength of localization is also determined by the detailed power spectrum of the potential and the energy of strongest localization depends on 
both the quasicondensate healing length $\xi$ and the disorder correlation length $\sigmar$.

Finally, let us discuss some possible extensions of our work.
On the experimental side, the observation of localized Bogoliubov excitations appears as a challenge.
It would be interesting to search for evidence of localization in the broadening of the dynamic structure factor, as measured in Bragg spectroscopy experiments~\cite{stenger1999,steinhauer2002,richard2003}.
Such broadenings have been measured to characterize coherence lengths of a couple hundred micrometers in elongated quasicondensates~\cite{richard2003}.
We infer therefrom that Bragg spectroscopy should allow the measurement of localization lengths of the same order of magnitude.
On the theoretical side, the localization properties of Bogoliubov quasiparticles in two and three dimensions are expected to exhibit an even richer phenomenology. In particular, as pointed out previously~\cite{lugan2010phd,gaul2011b}, the screening of disorder by interactions is expected to lead to the possible occurrence of two mobility edges in three dimensions. In such a scenario, a first delocalization transition would occur at high energy, as for bare particles, and a second one would occur at low energies, as the effect of disorder is suppressed in the limit of vanishing quasiparticle energy~\cite{john1983}. The localized states would then reside around $k\xi\simeq1$. Below a critical amount of disorder however, no quasiparticle states should be localized at all. As the screened potential derived here accurately describes scattering in higher dimensions as well, it may offer a simple avenue for the description of such a phenomenology.

%%%%%%%%%%%%%%%%%%%%%%%%%%%%%%%%%%%%%%%%%%%
\acknowledgements
We thank
A.~Aspect, P.~Bouyer, D.~Cl\'ement, T.-L.~Dao, L.~Fontanesi, C.~Gaul, V.~Gurarie, C.~A.~M\"uller, V.~Savona, and M.~Wouters for fruitful discussions.
This research was supported by
CNRS,
the European Research Council (FP7/2007-2013 Grant Agreement No.\ 256294),
Agence Nationale de la Recherche (Contract No.\ ANR-08-blan-0016-01),
RTRA-Triangle de la Physique, and the Alexander von Humboldt Foundation (Fellowship No.\ 1139948).
We acknowledge the use of the computing facility cluster GMPCS of the 
LUMAT federation (FR LUMAT 2764).
Laboratoire Charles Fabry de l'Institut d'Optique (LCFIO) is
a member of the Institut Francilien de Recherche sur les Atomes Froids (IFRAF).

%%%%%%%%%%%%%%%%%%%%%%%%%%%%%%%%%%%%%%%%%%%
\section*{APPENDICES}
\begin{appendix}

%%%%%%%%%%%%%%%%%%%%%%%%%%%%%%%%%%%%%%%%%%%
\section{Perturbation series of the smoothing solution}
\label{sec:smoothing2}

In this appendix, we calculate explicitly the leading terms of the modulations $\tV(\vectr)=\gnBm-g\nBEC(\vectr)$ of the mean-field interaction term $g\nBEC(\vectr)$ and the deviation $\Delta=\gnBm-\mu$ from the homogeneous equation of state. We consider the limit of a weak external potential $V$ and strong interactions which is relevant to our study (see Sec.~\ref{subsec:BECdensity}). Working along the lines of Ref.~\cite{lsp2006}, the weakness of $V$ suggests an expansion of the square root of the density in powers of $\Vr/\mu$, which we write as
%+++++++++++++++++++++++++++++++++++++++++%
\begin{equation}
\sqrt{\nBEC (\vectr)} = \sqrt{\frac{\mu}{g}}\ \left[\phi^{(0)} (\vectr) + \phi^{(1)} (\vectr) + \phi^{(2)} (\vectr) + \ldots \right],
\label{eq:develsqrtnBEC}
\end{equation}
%+++++++++++++++++++++++++++++++++++++++++%
where $\phi^{(0)} (\vectr) = 1$ is the solution in the absence of disorder \cite{pitaevskii2004}
and the functions $\phi^{(n)} (\vectr)$ are real-valued.
We thus have
%+++++++++++++++++++++++++++++++++++++++++%
\begin{equation}
\nBEC (\vectr) = \frac{\mu}{g}\ \sum_{i,j} \phi^{(i)} (\vectr) \phi^{(j)} (\vectr)
\label{eq:develsqrtnBECbis}
\end{equation}
%+++++++++++++++++++++++++++++++++++++++++%
and the quantities of interest at any order in the expansion series are readily obtained by collecting the terms at the corresponding order:
%+++++++++++++++++++++++++++++++++++++++++%
\begin{eqnarray}
\Delta^{(0)} &=& 0
\label{eq:Delta0} \\
\Delta^{(n)} &=& \mu\ \mathop{\sum_{0\leq i,j\leq n}}_{i+j=n} \mean{ \phi^{(i)} \phi^{(j)}},
\qquad \textrm{for } n \geq 1
\label{eq:Deltan}
\end{eqnarray}
%+++++++++++++++++++++++++++++++++++++++++%
and
%+++++++++++++++++++++++++++++++++++++++++%
\begin{eqnarray}
\widetilde{V}^{(0)} (\vectr) &=& 0
\label{eq:tV0} \\
\widetilde{V}^{(n)} (\vectr) &=& \Delta^{(n)} - \mu \mathop{\sum_{0\leq i,j\leq n}}_{i+j=n}
\phi^{(i)} (\vectr) \phi^{(j)} (\vectr),
\label{eq:tVn} \\
&& \textrm{for }n \geq 1.
\nonumber
\end{eqnarray}
%+++++++++++++++++++++++++++++++++++++++++%
The functions $\phi^{(n)} (\vectr)$ are determined by inserting the perturbation
series~(\ref{eq:develsqrtnBEC}) into the GPE~(\ref{eq:GPE}), which is equivalently written as
%+++++++++++++++++++++++++++++++++++++++++%
\begin{equation}
\left[-2\xi^2\vect{\nabla}^2 - 1 + \frac{V(\vectr)}{\mu} + \frac{g \nBEC (\vectr)}{\mu} \right] \sqrt{\nBEC(\vectr)} = 0,
\label{eq:GPEbis}
\end{equation}
%+++++++++++++++++++++++++++++++++++++++++%
and by collecting the terms of order $n$.
The explicit calculation of $\phi^{(n)}$ becomes increasingly tedious as the number of terms involved grows like $n^2$. Yet, the above perturbation hierarchy produces a simple recursion formula which can be used in analytical or numerical calculations.
Following the procedure outlined above, for all $n\geq1$, we obtain
%+++++++++++++++++++++++++++++++++++++++++%
\begin{equation}
 \phi^{(n)}=-\frac{1}{2}G_{\xi} * \left[ \frac{V}{\mu}\phi^{(n-1)}+\mathop{\sum_{i+j+k=n}}_{0\leq i,j,k\leq n-1}\phi^{(i)}\phi^{(j)}\phi^{(k)}\right],
\label{eq:phin}
\end{equation}
%+++++++++++++++++++++++++++++++++++++++++%
where $G_{\xi} (\vectr)$ is the Green function associated with the operator
$-\xi^2\vect{\nabla}^2 + 1$, which is best written in Fourier space as
%+++++++++++++++++++++++++++++++++++++++++%
\begin{equation}
\ftdim{G}_{\xi} (\vect{q})=\frac{(2\pi)^{-d/2}}{1+(|\vect{q}|\xi)^2},
\label{eq:Gxibis}
\end{equation}
%+++++++++++++++++++++++++++++++++++++++++%
and the convolution product is defined as
%+++++++++++++++++++++++++++++++++++++++++%
\begin{equation}
 (f*g)(\vectr) = \int \diffd \vectr'\ f(\vectr-\vectr') g(\vectr').
\end{equation}
%+++++++++++++++++++++++++++++++++++++++++%

Applying the recursive procedure up to second order, we find
%+++++++++++++++++++++++++++++++++++++++++%
\begin{eqnarray}
 \phi^{(0)} &=& 1 \label{eq:phi0}\\
 \phi^{(1)} &=& -\frac{1}{2\mu} G_{\xi} * V \label{eq:phi1}\\
 \phi^{(2)} &=& \frac{1}{4\mu^2}  G_{\xi} * \left[
	V (G_{\xi} * V) - \frac{3}{2} (G_{\xi} * V)^2
 \right]. \label{eq:phi2}
\end{eqnarray}
%+++++++++++++++++++++++++++++++++++++++++%
Then, using Eqs.~(\ref{eq:Deltan}) and (\ref{eq:tVn}), we find
%+++++++++++++++++++++++++++++++++++++++++%
\begin{eqnarray}
 \Deltaone &=& 0\\
\nonumber\\
 \tVone &=& G_{\xi} * V \nonumber\\
        &=&  \int \diffd\vectr'\ G_{\xi}(\vectr-\vectr') V(\vectr'), \label{eq:tVonebis}
\end{eqnarray}
%+++++++++++++++++++++++++++++++++++++++++%
and
%+++++++++++++++++++++++++++++++++++++++++%
\begin{eqnarray}
\Deltatwo &=& \frac{1}{4\mu} \left(
 \overline{G_{\xi} * \left[2 V (G_{\xi}*V) - 3 (G_{\xi}*V)^2\right]} \right.\nonumber\\
 && \left. \qquad \overline{+ (G_{\xi} * V)^2} \right) \nonumber\\
 &=& \frac{\Vr^2\sigmar^d}{2(2\pi)^{d/2}\mu}
            \int \diffd\vect{q}\ \frac{(|\vect{q}|\xi)^2}{\left[1+(|\vect{q}|\xi)^2\right]^2}\ctwoFT(\vect{q}\sigmar)
\label{eq:Deltatwobis}\\
\nonumber\\
\nonumber\\
\tVtwo &=& \Deltatwo - \frac{1}{4\mu} \left\{
 G_{\xi} * \left[2 V (G_{\xi}*V) - 3 (G_{\xi}*V)^2\right] \right.\nonumber\\
 && \left. + (G_{\xi} * V)^2 \right\}. \label{eq:tVtwobis}
\end{eqnarray}
%+++++++++++++++++++++++++++++++++++++++++%

Finally, let us make two comments on the above perturbative solution of the GPE.
First, for the perturbation expansion be valid, the mean-field density profile $\nBEC$
must be weakly perturbed around the homogeneous value $\mu/g$.
While the original small parameter is $\Vr/\mu$,
Eq.~(\ref{eq:phin}) shows that
%+++++++++++++++++++++++++++++++++++++++++%
\begin{equation}
\vert\tVoner\vert \ll \mu
\label{eq:weakLooser}
\end{equation}
%+++++++++++++++++++++++++++++++++++++++++%
with
%+++++++++++++++++++++++++++++++++++++++++%
\begin{equation}
\tVoner = \textrm{sign}(\Vr) \sqrt{\mean{\tV{{}^{(1)}}^2}}
\label{eq:tVonerbis}
\end{equation}
%+++++++++++++++++++++++++++++++++++++++++%
is a somewhat looser criterion for the successive terms of the expansion to be small.
The r.m.s.\ amplitude $\vert\tVoner\vert$ [see Eq.~(\ref{eq:tVonerbis})]
can be calculated explicitly from Eq.~(\ref{eq:tVonebis}):
%+++++++++++++++++++++++++++++++++++++++++%
\begin{equation}
\vert \tVoner \vert = \sqrt{\frac{\Vr^2\sigmar^d}{(2\pi)^{d/2}} \int\diffd\vect{q}\frac{\ctwoFT(\vect{q}\sigmar)}{[1+(|\vect{q}|\xi)^2]^2}}.
\end{equation}
%+++++++++++++++++++++++++++++++++++++++++%
Second, Eq.~(\ref{eq:phin}) can be used to show by induction that, whenever the potential $V$ is extended, then so are all the perturbation orders $\phi^{(n)}$ and $\tV^{(n)}$. In the disordered case, these simple considerations show how localization can be destroyed in a regime of weak interactions compatible with the mean-field approach.

%%%%%%%%%%%%%%%%%%%%%%%%%%%%%%%%%%%%%%%%%%%
\section{Decoupling basis for Bogoliubov-de Gennes equations in weak potentials}
\label{sec:decouplingBasis}

In this appendix, we motivate the introduction of the functions $\gp$ and $\gm$ (see Sec.~\ref{subsec:decouplingBdGEs}) to solve the BdGEs~(\ref{eq:degennes1}) and (\ref{eq:degennes2}), and justify the use of the Schr\"{o}dinger-like equation (\ref{eq:schroelike}) for weak potentials. Throughout the paper, we assume that the system lies in a box whose dimensions eventually tend to infinity to emulate the continuum limit,
and we impose periodic or homogeneous Dirichlet boundary conditions on the functions $\fp$ and $\fm$. In the latter case, the density $\nBEC$ and the BQP components $\fp$ and $\fm$ vanish at the system boundaries. However, in the limit $\xi/L\to 0$, where $L$ is the system size, and the absence of an external potential, the system can be regarded as homogeneous. Together with these boundary conditions, the BdGEs form the differential problem to be solved. The system of coupled equations~(\ref{eq:BdGEtV1}) and (\ref{eq:BdGEtV2}) can be rewritten as a differential problem in matrix form:
%+++++++++++++++++++++++++++++++++++++++++%
\begin{equation}
\label{eq:dzf}
 \xi^2 \nabla^2 \vectf(\vectr) = H_\epsilon(\vectr) \vectf(\vectr),
\end{equation}
%+++++++++++++++++++++++++++++++++++++++++%
where
%+++++++++++++++++++++++++++++++++++++++++%
\begin{equation}
\vectf (\vectr)=
\left(
\begin{array}{c}
 \fp (\vectr) \\
 \fm (\vectr)
\end{array}
\right),
\end{equation}
%+++++++++++++++++++++++++++++++++++++++++%
and $H_\epsilon (\vectr)= H_\epsilon^{(0)} + W(\vectr)$ is a real-valued, symmetric matrix, which depends on the energy $\epsilon$ and the position $\vectr$, with\footnote{In an equivalent formulation, $H_\epsilon$ can be regarded as an operator acting on two-vectors of functions $\fp$ and $\fm$. The position representation of Eq.~(\ref{eq:dzf}) is adopted here for simplicity.}
%+++++++++++++++++++++++++++++++++++++++++%
\begin{eqnarray}
&& H_\epsilon^{(0)} =
 \begin{pmatrix}
  0      & - \epsilon/2\mu \\
  - \epsilon/2\mu          & 1
 \end{pmatrix}
\label{eq:Hepsilon0} \\
\nonumber\\
&& W(\vectr) =
\frac{1}{2\mu}
 \begin{pmatrix}
  V + \Delta - \tV      & 0 \\
  0          & V + 3\Delta - 3\tV
 \end{pmatrix},
\label{eq:Hepsilon1}
\end{eqnarray}
%+++++++++++++++++++++++++++++++++++++++++%
where the position dependence of $V$ and $\tV$ was dropped for conciseness.
The two differential equations on $\fp$ and $\fm$ associated with Eq.~(\ref{eq:dzf}) are strongly coupled via the off-diagonal terms in $H_\epsilon^{(0)}$. Since $W$ is small (at most of first order in $\Vr$), it is worth working in the basis that diagonalizes $H_\epsilon^{(0)}$. Indeed, although the change of basis may introduce off-diagonal
(coupling) terms in $W$, these terms will remain small. We will then show that this approach is suitable for the set-up of a perturbation expansion.

%%%%%%%%%%%%%%%%%%%%%%%%%%%%%%%%%%%%%%%%%%%
\subsection{Bogoliubov-de Gennes equations in the decoupling basis}

In the absence of an external potential ($V=0$), the matrix $W$ vanishes identically. Then, the matrix $H_\epsilon = H_\epsilon^{(0)}$ has two eigenvalues,
%+++++++++++++++++++++++++++++++++++++++++%
\begin{eqnarray}
 \frac{1-\sqrt{1+(\epsilon/\mu)^2}}{2} & \equiv & -k^2 \xi^2 \label{eq:ke}\\
 \frac{1+\sqrt{1+(\epsilon/\mu)^2}}{2}  & \equiv & +\beta^2 \xi^2 \label{eq:betae},
\end{eqnarray}
%+++++++++++++++++++++++++++++++++++++++++%
associated to the eigenvectors
%+++++++++++++++++++++++++++++++++++++++++%
\begin{equation}
\label{eq:FkFbeta}
F_k \propto \left(
\begin{array}{c}
 \sqrt{\rho} \\
 +1/\sqrt{\rho}
\end{array}
\right)
~~~ \textrm{and} ~~~
F_\beta \propto \left(
\begin{array}{c}
 -1/\sqrt{\rho} \\
 \sqrt{\rho}
\end{array}
\right),
\end{equation}
%+++++++++++++++++++++++++++++++++++++++++%
respectively, where
%+++++++++++++++++++++++++++++++++++++++++%
\begin{equation}
 \rho = \frac{\mu}{\epsilon}+\sqrt{1+\left(\frac{\mu}{\epsilon}\right)^2}.
\label{eq:rhoAppendix}
\end{equation}
%+++++++++++++++++++++++++++++++++++++++++%

For simplicity, let us restrict our discussion to the 1D case.\footnote{The conclusions are naturally extended in higher dimensions.} Since Eq.~(\ref{eq:dzf}) is of second order, each eigen-subspace corresponds to two possible solutions of the BdGEs.
First, the solutions corresponding to $+\beta^2\xi^2$ are $e^{\pm \beta z}F_\beta$, which either grow or decrease exponentially. These modes are the \textit{evanescent} modes discussed below Eq.~(\ref{eq:beta}), which thus appear naturally in this formulation. Since no solution of Eq.~(\ref{eq:dzf}) in the subspace spanned by these eigenvectors can satisfy the boundary conditions on both boundaries, these modes are forbidden in the case of vanishing $V$. Second, the solutions corresponding to $-k^2\xi^2$ are $e^{\pm\i k z}F_k$, which are \textit{oscillating} (plane wave) modes.
Since solutions of Eq.~(\ref{eq:dzf}) that satisfy the boundary conditions can be built
with linear combinations of these eigenvectors, these modes are allowed for $V=0$. They correspond to the well-known physical solutions of the BdGEs~(\ref{eq:degennes1}) and (\ref{eq:degennes2}).

The procedure above allowed us to decouple the BdGEs in homogeneous space. Let us introduce now the external potential $V(\vectr)$. Rewriting Eq.~(\ref{eq:dzf}) in the eigenbasis of $H_\epsilon^{(0)}$, we find
%+++++++++++++++++++++++++++++++++++++++++%
\begin{equation}
\label{eq:dzg}
 \xi^2 \nabla^2 \vectg(\vectr) = H_\epsilon^\prime (\vectr) \vectg(\vectr),
\end{equation}
%+++++++++++++++++++++++++++++++++++++++++%
where $\vectg(\vectr)=P^{-1}\vectf(\vectr)$, and
%+++++++++++++++++++++++++++++++++++++++++%
\begin{equation}
P^{-1} =
\left(
\begin{array}{c c}
 +\sqrt{\rho} & +1/\sqrt{\rho} \\
 -1/\sqrt{\rho} & +\sqrt{\rho}
\end{array}
\right)
\end{equation}
%+++++++++++++++++++++++++++++++++++++++++%
is the inverse of the transformation matrix from the $F$-basis to the $G$-basis. The term ${H}_\epsilon^\prime = {H}_\epsilon^\prime{}^{(0)} + W'$, which is the analog of $H_\epsilon$ in the $G$-basis, contains a homogeneous part
%+++++++++++++++++++++++++++++++++++++++++%
\begin{eqnarray}
 {H}_\epsilon^\prime{}^{(0)} &=&
 \begin{pmatrix}
  - k^2\xi^2    & 0 \\
     0          & \beta^2\xi^2
 \end{pmatrix},
\label{eq:HepsilonP0}
\end{eqnarray}
%+++++++++++++++++++++++++++++++++++++++++%
and a potential-dependent part
%+++++++++++++++++++++++++++++++++++++++++%
\begin{eqnarray}
{W'}(\vectr) &=& P^{-1} {W}(\vectr) P 
\label{eq:HepsilonP1} \\
\nonumber \\
 &=& \frac{1}{2\mu}
\left(
\begin{array}{c c c}
 V - \frac{3+\rho^2}{1+\rho^2} (\tV-\Delta) && -\frac{2\rho}{1+\rho^2} (\tV-\Delta) \\
&& \\
 V -\frac{2\rho}{1+\rho^2} (\tV-\Delta) && - \frac{1+3\rho^2}{1+\rho^2} (\tV-\Delta)
\end{array}
\right).
\nonumber
\end{eqnarray}
%+++++++++++++++++++++++++++++++++++++++++%
Note that the BQP energy $\epsilon$ is here embedded in the dependence of $k$, $\beta$ and $\rho$ on $\epsilon$ [see Eqs.~(\ref{eq:ke}), (\ref{eq:betae}) and~(\ref{eq:rhoAppendix})].

For the sake of clarity, let us write explicitly the two coupled equations associated to Eq.~(\ref{eq:dzg}) in terms of the components $\gpm$ of $\vectg (\vectr) = \left(\gp (\vectr),\gm (\vectr)\right)^\textrm{T}$:
%+++++++++++++++++++++++++++++++++++++++++%
\begin{eqnarray}
\frac{\hbar^2 k^2}{2m} \gp
& = &
- \frac{\hbar^2}{2m} \nabla^2 \gp
- \frac{2\rho}{1+\rho^2} (\tV-\Delta) \gm
\nonumber \\
&& + \left[V-\frac{3+\rho^2}{1+\rho^2} (\tV-\Delta)\right] \gp
\label{eq:decoupled1Appendix} \\
- \frac{\hbar^2 \beta^2}{2m} \gm
& = &
- \frac{\hbar^2}{2m} \nabla^2 \gm
- \frac{2\rho}{1+\rho^2} (\tV-\Delta) \gp
\nonumber \\
&&  + \left[V-\frac{1+3\rho^2}{1+\rho^2} (\tV-\Delta)\right] \gm,
\label{eq:decoupled2Appendix}
\end{eqnarray}
%+++++++++++++++++++++++++++++++++++++++++%
where
%+++++++++++++++++++++++++++++++++++++++++%
\begin{equation}
\label{eq:gpmbis}
g^\pm_{\nu} (\vectr) = \pm\rho_{\nu}^{\pm 1/2}\fp_{\nu}(\vectr) +\rho_{\nu}^{\mp 1/2} \fm_{\nu}(\vectr).
\end{equation}
%+++++++++++++++++++++++++++++++++++++++++%
These equations are equivalent to the BdGEs, without any approximation. The benefit of the transformation we have used is that, for a weak external potential, the terms appearing in ${W}'$ are all small. In particular, the terms coupling $g^+$ and $g^-$ in Eqs.~(\ref{eq:decoupled1Appendix}) and~(\ref{eq:decoupled2Appendix}) are at most of the order of $V$
(as $\vert\tVr\vert \simeq \vert\tVoner\vert \le \vert\Vr\vert$ and $2\rho/(1+\rho^2) \leq 1$). The $G$-basis thus offers a suitable starting point, which takes into account the full structure of the BdGEs, and which allows for a perturbative approach in the regime of weak disorder.

%%%%%%%%%%%%%%%%%%%%%%%%%%%%%%%%%%%%%%%%%%%
\subsection{Leading-order terms: mapping the Bogoliubov-de Gennes equations onto a Schr\"{o}dinger-like equation}
\label{subsec:mappingAppendix}

\begin{figure}[t!]
\begin{center}
\includegraphics[width=8cm]{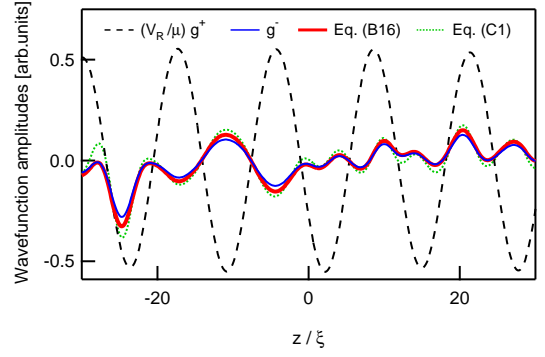}
\end{center}
\caption{(color online) BQP mode in the $\gpm$ basis. This eigenmode was computed at energy $\epsilon\simeq1.1\mu$ for a speckle potential with $\Vr=0.05\mu$ and $\sigmar/\xi=\sqrt{3/2}$. The mode is displayed here over a few healing lengths, while the  size of the box used for the calculation is $L= 3.2\times10^5 \xi$. The $\gp$ and $\gm$ components obtained numerically are given by the black dashed line and the blue solid line, respectively. The thick red line represents the convolution formula~(\ref{eq:gkmConvolution}), and the green dotted line the somewhat cruder approximation~(\ref{eq:approxtVgm}) to $\gm$. Note that the $\gp$ component has been rescaled by $\Vr/\mu$ in this figure for better comparison with the various representations of $\gm$. With the above parameters, $\tVoner\simeq0.6\,\Vr$.
}
\label{fig:gkmConvolution}
\end{figure}

Let us now develop the perturbation expansion of the BdGEs in the $G$-basis. Since Eqs.~(\ref{eq:decoupled1Appendix}) and (\ref{eq:decoupled2Appendix}) are weakly coupled, we can resort to the following self-consistent approach. Assuming that $\gm$ is vanishingly small compared to $\gp$ for small $\Vr$, we neglect the third term on the right-hand side of Eq.~(\ref{eq:decoupled2Appendix}). Then, solving for $\gm$ and retaining only the leading-order term in $\Vr$, we obtain\footnote{Note that $\Delta$ is of second order in $\Vr$ (see Sec.~\ref{subsec:BECdensity} or Appendix~\ref{sec:smoothing2}).}
%+++++++++++++++++++++++++++++++++++++++++%
\begin{equation}
\label{eq:gkmConvolution}
\gm (\vectr) \simeq
\frac{2m}{\hbar^2\beta^2}\frac{2\rho}{1+\rho^2} \int \diffd\vectr'
G_{1/\beta}(\vectr-\vectr') \tVone (\vectr') \gp (\vectr'),
\end{equation}
%+++++++++++++++++++++++++++++++++++++++++%
where $G_{1/\beta}(\vect{q})=\frac{(2\pi)^{-d/2}}{1+(|\vect{q}|/\beta)^2}$ is the Green function associated with the differential operator $-(1/\beta)^{2}\nabla^2+1$, written in Fourier space. The positive smoothing function $G_{1/\beta}$ satisfies $\int \diffd\vectr' G_{1/\beta}(\vectr') = 1$, and decays on the length scale $1/\beta$, which is smaller than the healing length $\xi$ and than $1/k$, \ie the typical length scale over which $\gp$ varies (see Fig.~\ref{fig:gkmConvolution}). Thus, owing to the fact that $2m/(\hbar^2\beta^2)<1/\mu$ and $2\rho/(1+\rho^2)<1$, 
we can safely write
%+++++++++++++++++++++++++++++++++++++++++%
\begin{equation}
|\gm(\vectr)|  <  \frac{1}{\mu} \int \diffd\vectr' G_{1/\beta}(\vectr-\vectr') |\tVone(\vectr')| |\gp(\vectr')|.
\end{equation}
%+++++++++++++++++++++++++++++++++++++++++%
Then, in terms of orders of magnitude,
%+++++++++++++++++++++++++++++++++++++++++%
\begin{eqnarray}
\label{eq:gkmBound}
|\gm|
&\lesssim& \frac{\vert\tVoner\vert}{\mu} |\gp|
\int \diffd\vectr' G_{1/\beta}(\vectr-\vectr') \nonumber \\
&\lesssim& \frac{\vert\tVoner\vert}{\mu} |\gp| \ll |\gp|,
\end{eqnarray}
%+++++++++++++++++++++++++++++++++++++++++%
which is consistent with our initial assumption, \ie $\gm$ is small compared
to $\gp$.
In Fig.~\ref{fig:gkmConvolution} we show numerical results which corroborate expression (\ref{eq:gkmConvolution}) and the fact that $\gm$ is a term of order $\tVoner/\mu$ at most to $\gp$.
%-----------------------------------------%
%-----------------------------------------%

From the upper bound (\ref{eq:gkmBound}), we infer that the second term on the right hand side of Eq.~(\ref{eq:decoupled1Appendix}) is of the order of $(\tVr^2/\mu)|\gp|$ at most, while the third term contains terms scaling as $|\Vr\gp|$. Hence, we neglect the former contribution, and obtain a closed equation for $\gp$ which is valid up to first order in $\Vr/\mu$:
%+++++++++++++++++++++++++++++++++++++++++%
\begin{equation}
-\frac{\hbar^2}{2m} \nabla^2 \gp +\mVk (\vectr) \gp 
\simeq \frac{\hbar^2 k^2}{2m} \gp,
\label{eq:schroelikeAppendix} \\
\end{equation}
%+++++++++++++++++++++++++++++++++++++++++%
where
%+++++++++++++++++++++++++++++++++++++++++%
\begin{equation}
\mVk (\vectr) = V(\vectr) - \frac{3 + \rho^2}{1 + \rho^2} \tVone (\vectr).
\label{eq:effectiveVAppendix}
\end{equation}
%+++++++++++++++++++++++++++++++++++++++++%
These are the expressions reproduced in Eqs.~(\ref{eq:schroelike}) and (\ref{eq:effectiveV}), which form the basis of our approach to calculate the BQP modes in leading-order perturbation theory. The advantage of the $\gpm$ representation is that, at this level of approximation, the coupled equations (\ref{eq:BdGEtV1}) and (\ref{eq:BdGEtV2}) reduce to a simple closed equation for $\gp$, the solution of which also determines $\gm$ via Eq.~(\ref{eq:gkmConvolution}).

%%%%%%%%%%%%%%%%%%%%%%%%%%%%%%%%%%%%%%%%%%%
\section{Lyapunov exponent of Bogoliubov quasiparticles beyond the Born approximation}
\label{sec:appendixBeyondBorn}

In this appendix, the perturbation expansion of the Lyapunov exponent of BQPs is extended one order beyond the leading-order (Born) approximation, so as to include terms scaling as $\Vr^3$. To this aim, we consider explicitly the terms in Eq.~(\ref{eq:decoupled1Appendix}) that are of second order in $\Vr$. 

The third term on the right hand side of Eq.~(\ref{eq:decoupled1Appendix}) contains both an inhomogeneous term $\tVtwo$ and an offset $\Deltatwo$ that are proportional to $\Vr^2$, and for which explicit expressions are given in appendix \ref{sec:smoothing2}. Elements of order $\Vr^2$ are also introduced into Eq.~(\ref{eq:decoupled1Appendix}) by the cross-term $\tV \gm$, where $\gm$ may be replaced by expression~(\ref{eq:gkmConvolution}).
If $\gp$ varies on a length scale larger than that of the other quantities in the integrand of Eq.~(\ref{eq:gkmConvolution}), we can use the approximation
%+++++++++++++++++++++++++++++++++++++++++%
\begin{equation}
\label{eq:approxtVgm}
%|g_\vect{k}^-/g_\vect{k}^+| \lesssim 
%\frac{2/\epsk}{1+\rho_k^2} |\tV| < |\tV|/\mu \ll 1,
\gm (\vectr) \simeq
\frac{2m}{\hbar^2\beta^2}\frac{2\rho}{1+\rho^2}  \gp (\vectr) \int \diffd\vectr'
G_{1/\beta}(\vectr-\vectr') \tV (\vectr').
\end{equation}
%+++++++++++++++++++++++++++++++++++++++++%
While Eq.~(\ref{eq:approxtVgm}) is justified for $k \ll \mathrm{min}(1/\sigmar,1/\xi)$, we found that it is actually a good approximation on a broader range of parameters (see Sec.~\ref{subsec:beyondBorn}).
For instance, this approximation shows good agreement with direct numerical results for $\gm$ for the parameters of Fig.~\ref{fig:gkmConvolution}.
Hence, we get a new closed equation for $\gp$, which now comprises all the terms up to order $\Vr^2$ and is legitimate in the low-energy limit:
%+++++++++++++++++++++++++++++++++++++++++%
\begin{equation}
\frac{\hbar^2 k^2}{2m} \gp
\simeq
- \frac{\hbar^2}{2m} \nabla^2 \gp
+\left[\mVk(\vectr)+\mVn(\vectr)+\mVm(\vectr) \right] \gp,
\label{eq:decoupledSecondOrderAppendix}
\end{equation}
%+++++++++++++++++++++++++++++++++++++++++%
where $\mVk$ is the screened potential~(\ref{eq:effectiveVAppendix}),
and the terms $\mVn$ and $\mVm$ are potentials proportional to $\Vr^2$:
%+++++++++++++++++++++++++++++++++++++++++%
\begin{eqnarray}
\mVn(\vectr) & = & -\frac{3+\rho^2}{1+\rho^2} \left[\tVtwo(\vectr)-\Deltatwo \right] \label{eq:mVn}\\
\mVm(\vectr) & = & - \frac{8m\rho^2}{\hbar^2\beta^2\left(1+\rho^2\right)^2} \times \nonumber \\ 
&&\int \diffd\vectr' G_{1/\beta}(\vectr-\vectr') \tVone(\vectr)\tVone (\vectr'). \label{eq:mVm}
\end{eqnarray}
%+++++++++++++++++++++++++++++++++++++++++%
The potential term $\mVn$ follows from a second-order expansion of the ground-state density profile, and $\mVm$ originates from the coupling between $\gp$ and $\gm$. Both $\mVn$ and $\mVm$ have a non-vanishing average. These non-vanishing averages suggest an evaluation of the correlation functions at a wave vector off the energy shell~(\ref{eq:BogoliubovDispersion}) in the fourth-order Lyapunov exponent $\Gamma^{(4)}$. However, these averages play no role in the correlation functions contributing to $\Gamma_k^{(3)}$ (see below), and can thus be disregarded at this level of approximation.

For Schr\"odinger particles of energy $E=\hbar^2 k^2/2m$ in a 1D disordered potential $V$, the leading orders of the weak-disorder expansion of the Lyapunov exponent read~\cite{lugan2009}:
%+++++++++++++++++++++++++++++++++++++++++%
\begin{equation}
\gamma_k^{(2)}=\frac{1}{4 k^2}\left(\frac{2m}{\hbar^2}\right)^2 \int_{-\infty}^{0} \diffd z \Ctwo(z)\cos(2k z),
\end{equation}
%+++++++++++++++++++++++++++++++++++++++++%
with $\Ctwo(z)=\mean{V(0)V(z)}$ and 
%+++++++++++++++++++++++++++++++++++++++++%
%\begin{widetext}
\begin{eqnarray}
\gamma_k^{(3)}&=&-\frac{1}{4k^3}\left(\frac{2m}{\hbar^2}\right)^3 \mathcal{P}\int \diffd q \frac{\CthreeFT(q,2k)+\CthreeFT(-q,-2k)}{2q}\nonumber\\
&=&-\frac{1}{4 k^3}\left(\frac{2m}{\hbar^2}\right)^3
\int_{-\infty}^{0} \diffd z \int_{-\infty}^{z} \diffd z^\prime \Cthree(z,z^\prime) \sin(2k z^\prime)\nonumber\\
\end{eqnarray}
%\end{widetext}
%+++++++++++++++++++++++++++++++++++++++++%
where $\mathcal{P}$ denotes a Cauchy principal value, $\Cthree(z,z^\prime)=\mean{V(0)V(z)V(z^\prime)}$ is the three-point correlation function and $\CthreeFT(q,q^\prime)$ is its Fourier transform. Replacing $V$ in these formulas by the sum of the potential terms appearing in Eq.~(\ref{eq:decoupledSecondOrderAppendix}) and collecting the different terms according to their order in $\Vr$, up to $\Vr^3$, we find
%+++++++++++++++++++++++++++++++++++++++++%
\begin{equation}
\Gamma_k \simeq
\Gamma^{(2)}_{\mVk,\mVk}+
\Gamma^{(3)}_{\mVk,\mVn}+
\Gamma^{(3)}_{\mVk,\mVm}+
\Gamma^{(3)}_{\mVk,\mVk,\mVk},
\end{equation}
%+++++++++++++++++++++++++++++++++++++++++%
where
%+++++++++++++++++++++++++++++++++++++++++%
\begin{equation}
\label{eq:contrib1}
 \Gamma^{(2)}_{\mVk,\mVk}  = 
\frac{1}{4 k^2}\left(\frac{2m}{\hbar^2}\right)^2 \int_{-\infty}^{0}\diffd z \mean{\mVk(0)\mVk(z)} \cos(2 {k} z)
\end{equation}
%+++++++++++++++++++++++++++++++++++++++++%
corresponds to result~(\ref{eq:Gk2}), obtained in the Born approximation, and the remaining terms are the three contributions to the third-order Lyapunov exponent of BQPs:
%+++++++++++++++++++++++++++++++++++++++++%
\begin{widetext}
\begin{eqnarray}
\label{eq:contrib2}
\Gamma^{(3)}_{\mVk,\mVn}&=&\frac{1}{4 k^2}\left(\frac{2m}{\hbar^2}\right)^2 \int_{-\infty}^{0}\diffd z \big[\mean{\mVk(0)\mVn(z)}+\mean{\mVn(0)\mVk(z)}\big] \cos(2 {k} z)\\
\label{eq:contrib3}
\Gamma^{(3)}_{\mVk,\mVm}&=&\frac{1}{4 k^2}\left(\frac{2m}{\hbar^2}\right)^2 \int_{-\infty}^{0}\diffd z \big[\mean{\mVk(0)\mVm(z)}+\mean{\mVm(0)\mVk(z)}\big] \cos(2 {k} z)\\
\label{eq:contrib4}
\Gamma^{(3)}_{\mVk,\mVk,\mVk}&=&-\frac{1}{4 k^3}\left(\frac{2m}{\hbar^2}\right)^3 \int_{-\infty}^{0} \diffd z \int_{-\infty}^{z} \diffd z^\prime \mean{\mVk(0)\mVk(z)\mVk(z^\prime)} \sin(2 {k} z^\prime).
\end{eqnarray}
\end{widetext}
%+++++++++++++++++++++++++++++++++++++++++%
Note that, since $\mean{\mVk}=0$, the non-vanishing mean values of $\mVn$ and $\mVm$ play no role in the various contributions to $\Gamma^{(3)}$, and can substracted from $\mVn$ and $\mVm$ in Eqs.~(\ref{eq:contrib2}) and~(\ref{eq:contrib4}). As the expressions of contributions~(\ref{eq:contrib2}) to~(\ref{eq:contrib4}) in Fourier space are quite involved, we do not reproduce them here. We refer to Fig.~\ref{fig:GammaAnalyticNumeric} and to section~\ref{subsec:beyondBorn}, which provide a discussion of the behavior of these terms.

\end{appendix}

%%%%%%%%%%%%%%%%%%%%%%%%%%%%%%%%%%%%%%%%%%%%%%%%%%%%%%%%%%%%%%%%%
\newcommand{\Jnature}{Nature (London)}
\newcommand{\Jnatphys}{Nature Phys.}

\newcommand{\Jprl}{Phys. Rev. Lett.}
\newcommand{\Jpr}{Phys. Rev.}
\newcommand{\Jpra}{Phys. Rev. A}
\newcommand{\Jprb}{Phys. Rev. B}
\newcommand{\Jprc}{Phys. Rev. C}
\newcommand{\Jprd}{Phys. Rev. D}
\newcommand{\Jpre}{Phys. Rev. E}
\newcommand{\Jrmp}{Rev. Mod. Phys.}

\newcommand{\Jepl}{Europhys. Lett.}
\newcommand{\Jnjp}{New J. Phys.}
\newcommand{\Jepjd}{Eur. Phys. J. D}

\newcommand{\Jjosab}{J. Opt. Soc. Am. B}
\newcommand{\JApplPhysLett}{Appl. Phys. Lett.}

\newcommand{\JJlowT}{J. Low Temp. Phys.}

\newcommand{\Jjetp}{Sov. Phys. JETP}
\newcommand{\JjphysUSSR}{J. Phys. USSR}

\newcommand{\Jjphyschemsol}{J. Phys. Chem. Sol.}
\newcommand{\Jsolstatecomm}{Solid State Comm.}

\newcommand{\Jijmpb}{Int. J. Mod. Phys. B}

\newcommand{\JphysicaBC}{Physica B+C}

\newcommand{\Jphysrep}{Phys. Rep.}
\newcommand{\JRepProgPhys}{Rep. Prog. Phys.}
\newcommand{\JjphysCM}{J. Phys.: Cond. Matt.}
\newcommand{\JjphysA}{J. Phys. A: Math. Gen.}
\newcommand{\Jphystoday}{Phys. Today}

\newcommand{\Jadvatmoloptphys}{Adv. At. Mol. Opt. Phys.}

\newcommand{\JphysB}{J. Phys. B: At. Mol. Opt. Phys.}

\newcommand{\JTheorMathPhys}{Theor. Math. Phys.}
\newcommand{\Jprogthphyssup}{Prog. Theor. Phys., Suppl.}

\newcommand{\JjphysF}{J. Phys. (France)}
\newcommand{\JjphysquatreF}{J. Phys. IV (France)}
% \bibliography{$HOME/Documents/work/publications/papers/bibliography/biblioLSP}
\bibliography{biblioLSP_20110429}

\end{document}